\documentclass[a4paper,10pt]{article}
\usepackage{fancyhdr}
\usepackage{fancybox, framed}
\usepackage{graphicx}
\usepackage{url}
\usepackage{amsbsy,amssymb}
\usepackage{amsmath}
\usepackage{abstract}
\usepackage{hyperref}
\usepackage{doi}
\usepackage{color}
\usepackage{authblk}
\usepackage{amsthm}
\usepackage{bm}

\usepackage{cite}
\usepackage{mhchem,chemformula}

\usepackage[margin=1.5cm,marginparwidth=2cm]{geometry}
\usepackage{marginnote}

\usepackage{fancyvrb}

\usepackage{pdflscape}
\usepackage{xcolor}
\usepackage{lipsum}
\usepackage{graphicx}
 \usepackage{lineno}

\usepackage[caption = false]{subfig}
\usepackage{lscape}
\usepackage[normalem]{ulem}

\edef\marginnotetextwidth{\the\textwidth}
\DeclareMathOperator{\tr}{tr}

\makeatletter
\renewcommand*{\raggedrightmarginnote}{\raggedright\if@firstcolumn\(\triangleleft \) \fi}
\makeatother

\title{Elastodynamical properties of Sturmian structured media}
\author[1,2]{M. L\'azaro\footnote{Corresponding author, mail: malana@upv.es}}
\author[3]{A. Niemczynowicz}
\author[3]{A. Siemaszko}
\author[1]{L.M. Garcia-Raffi}

\affil[1]{Instituto Universitario de Matem\'atica Pura y Aplicada, Universitat Polit\`ecnica de Val\`encia.  Spain}
\affil[2]{Department of Continuum Mechanics and Theory of Structures, Universitat Polit\`ecnica de Val\`encia.  Spain}
\affil[3]{Faculty of Mathematics and Computer Science, University of Warmia and Mazury in Olsztyn, Poland}

\begin{document}
\maketitle

\begin{abstract}
In this paper, wave propagation in structured media with quasiperiodic patterns is investigated. We propose a methodology based on Sturmian sequences for the generation of structured mechanical systems from a given parameter. The approach is presented in a general form so that it can be applied to waveguides of different nature, as long as they can be modeled with the transfer matrix method. The bulk spectrum is obtained and its fractal nature analyzed. For validation of the theoretical results, three numerical examples are presented. The obtained bulk spectra show different shapes for the studied examples, but they share features which can be explained from the proposed theoretical setting. 
\end{abstract}

\section{Introduction}


The guiding and manipulation of waves is a major issue in Science and Technology. In the last decades, the manipulation of waves and scattering have got rapid progress with the development of artificial materials known as photonic, phononic and sonic crystals~\cite{Joannopoulos-2011,Sigalas-1992,Deymier-2013}. They are periodic structures built by repetition of unit cell in one or several dimensions. The particular design of the unit cell allows customization of the wave behavior by the appearance of passbands and stopbands in the frequency domain. A frequency is said to be forbidden (or it lies in a bandgap-stopband) if the associated wave is evanescent with amplitudes decaying along the structure. The final objective is to tailor the dispersion of waves enabling a precise control over the spectral transmission, velocity and dynamic properties.\\

Another possibility to manipulate waves in this respect is the use of quasiperiodic systems.. They are characterized by deterministic patterns that exhibit correlation in the  long-range order. Therefore, quasiperiodic media present unique properties that have been the subject of extensive research in various areas of physics: electronics~\cite{Steinhardt-1987}, electromagnetis~\cite{Man-2005}, elasticity~\cite{Liu-2004,Ruzzene-2019}. One of the most visible features of quasiperiodicity is the self-similarity of the spectrum of permitted modes and frequencies~\cite{Macia-2009,Velasco-2001}. In particular, this behavior has been observed in systems constructed with the Fibonacci formalism~\cite{Kolar-1992,Lange-1981}.\\

 In this paper, Sturmian quasiperiodic lattices are under consideration. A Sturmian word (or sequence) is a particular case of an infinite word formed from a two-letter alphabet and, among other applications, allows the construction of quasiperiodic patterns. The precise definition and properties of Sturmian words can be found in the context of computer science and language theory~\cite{Fogg-2002} and condensed matter~\cite{Damanik-1999a,Damanik-1999b}. Although in essence, the concept allows several definitions~\cite{Crisp-1993,Lothaire-2002}, among them we are interested in such sequences that can be generated from real numbers (also called mechanical words~\cite{Berstel-1995}). \\


In the past two decades the number of works on the design of quasiperiodic structures has been proliferating. Such structures have many interesting properties that can be useful for practical applications in crystallography, photonics, or structural mechanics. Analysis of wave propagation in periodic and quasiperiodic structures is the main topic in many engineering areas. Dynamical analysis of such structures is critical because it allows us to improve vibroacoustic properties. A less investigated field in structural mechanics is that the design and dynamical analysis of quasiperiodic structures. Appropriate design of such quasiperiodic structures may leads to new interesting vibroacoustic properties with a wide range of applications. Among many possibilities of design of 1D quasiperiodic lattices, the most popular are the methods based on Fibonacci number sequences \cite{Merlin1985,Kohmoto1987,Engel2007}. The dispersion relation of these systems exhibit band gaps, which often take the form of the well-known Hofstadter butterfly \cite{Hofstadter-1976}. In mechanics, despite a few studies described below, the dispersion properties of quasiperiodic elastic media have not yet been satisfactorily understood, specially  the design of tailored structures with specific properties concerning wave propagation. Elastodynamical properties of finite or infinite periodic 1D rods or beams have been the topic of research investigations in  \cite{ Richards2003,Hussein2006,Yu2006, Liu2012}. Recent developments of the research in this field comprise the analysis of how localized modes arise in continuous elastic media with quasiperiodic stiffness modulation \cite{Ruzzene-2019} or the analysis of the effects of combined modulation of structural parameters, with different arbitrarily related spatial periods, on wave propagation properties of a general 1D waveguide \cite{Sorokin2019}. For example, in ref. \cite{Ruzzene-2020c} the topological modes for stiffened and sandwich beams are investigated. The authors demonstrated that the occurrence of bandgaps possesses fractal nature in the frequency spectrum of 1D continuous quasiperiodic elastic media (continuous stiffened beams and sandwich beams). Another interesting example of the analysis of the structural response of periodic and quasiperiodic beams is presented in ref. \cite{ Timorian2019}, wherein geometrical and material variations are introduced following Fibonacci patterns along the spans modeled by finite elements. Glacet et al. \cite{Glacet2019} provided the complete description of the vibrational analysis of a beam based on octagonal quasiperiodic tiling, while Gei \cite{Gei2010} studied the dispersion relations for axial and flexural waves of quasiperiodic infinite beams. Srivastava et al. \cite{Srivastava2010}  studied theoretically the occurrence of longitudinal or flexural waves but in nonlinear isotropic rods. \\

In the literature we can find different examples on the study of the spectrum of allowed modes as a function of a certain parameter that generates the quasiperiodic pattern. The most relevant case is that of the aforementioned Hoesteadtler butterfly~\cite{Hofstadter-1976} in condensed matter, which has been reproduced and studied in other works~\cite{Lange-1981,Fuchs2016}. Following the generation pattern known as the projection method~\cite{Elser} other authors have obtained similar figures in other physical systems like discrete mass-spring lattices~\cite{Silva-2019}, quasiperiodic beams~\cite{Ruzzene-2019}, dielectric quasicrystals~\cite{Rodriguez-2008}, acoustic metamaterials~\cite{Cheng-2020} or quasicrystals of magnetic resonators~\cite{Apigo-2018}. \\


In this paper we propose to investigate systems which are structured according to quasiperiodic patterns governed by the so-called Sturmian sequences, something that will be carried out in the context of structural dynamics. After a rigorous definition of structured systems based on Sturmian sequences, a geometrical interpretation is provided and a systematic methodology for the determination of dispersion relations, valid for unidirectional lattices, is defined. We also introduce the concept of  Sturmian bulk spectrum and study the observed self-similarity properties. Finally, the theoretical results are validated numerically through several numerical examples covering discrete and continuous systems, as well as compression and bending waves.  In this sense, the application of Sturmian sequences to mechanical engineering contributes to the design of mechanical structures with tailored properties for wave propagation.

\section{Sturmian quasiperiodic structures}

In this section, the construction of structural systems with quasiperiodic pattern based on Sturmian sequences will be presented. It will be considered that the wave is propagated in one direction throughout a infinite elastic medium. There is no restriction on the nature of the waves. Thus, as it will be seen in the numerical examples, different types of waves and structures can be considered. Elastic wave transmission models depend on several parameters, in general related to the geometry, mass and stiffness properties mostly. If for instance, we deal with a continuous homogeneous non-dispersive medium, then waves travel with constant velocity and all waves are permitted, that is for each frequency $\omega$, there exist a non-evanescent mode with wavenumber $\kappa(\omega)$. If, on the contrary, one of the parameter changes along the waveguide under certain pattern of periodicity, then the dispersion curves may present band gaps, where the wave is damped in the space with complex wavenumber. The way in which the parameters are distributed opens up a tremendous range in terms of the typology of systems obtained. Thus quasiperiodic patterns arise as an extension of periodic ones, although we can also find order in apparently random geometries, such as hyperuniformity~\cite{Torquato-2002}.

\subsection{Sturmian sequencies}

We will study the dynamical properties of elastic waves considering that certain model parameter follows the pattern of a Sturmian word along the medium. In our work, we are specially interested in the generation algorithm of a Sturmian word from a real number, which along this paper will be denoted by $\alpha \in \mathbb{R}$. This latter plays the roll of {\em generation parameter} and, without loss of generality, it lies in the range $0 \leq \alpha \leq 1$. Let the sequence $[0;a_1,\ldots,a_n]$ be the continuous fraction of $\alpha$, namely 
\begin{equation}
\alpha = [0;a_1,\ldots,a_n] =
\frac{1}{a_1 + \dfrac{\phantom{1}}{\cdots  + \dfrac{1}{a_{n-1} + \dfrac{1}{a_n}}} }  \ , 
\label{eq001}
\end{equation}
where $a_k > 0$, for $k \geq 1$, are positive integer numbers. Consider in addition a binary alphabet formed by two symbols, say $\{p,q\}$. Then, we define a Sturmian word in a recursively way as the sequence of symbols
\begin{eqnarray}
\mathcal{B}_k &=& 
\mathcal{B}_{k-1}^{a_k} \ 
\mathcal{B}_{k-2}  \ , \quad  1 \leq k \leq n \ , 
\nonumber \\
\mathcal{B}_{-1} &=& q \ , \quad \mathcal{B}_{0} = p \ , 
\label{eq002}
\end{eqnarray}
where both the exponent and the product must be understood as concatenations, for instance $p^3(q^2p) = pppqqp$. From the definition given above, if $\alpha$ is a rational number the Sturmian word $\mathcal{B}_n$ is properly the last iteration and strictly the infinite word arises as the periodic concatenation of $\mathcal{B}_n$. Thus, for instance if $\alpha = 1/3 = [0;3]$, then the last block is $\mathcal{B}_1 = pppq$ and the infinite word $pppq \, pppq \, pppq \, \ldots$. Otherwise, if $\alpha$ is irrational, then it is known that the sequence $a_n$ becomes infinite and the associated Sturmian word has a purely quasiperiodic pattern given by the limit $\lim_{n\to \infty} \mathcal{B}_n$. This form of constructing a Sturmian word is not unique, in fact there are other ways of finding Sturmian patterns~\cite{Berstel-1995,Crisp-1993}. Although they have different geometrical interpretations and recursive models, still they are closely related each other~\cite{Lothaire-2002}. In particular, it will be shown later in Sec.~\ref{GeometricalInterpretation} that Sturmian words defined as in Eq.~\eqref{eq002} have an interesting geometrical interpretation. \\

Each one of the words emerging from the recursive sequence \eqref{eq002} will be named {\em Sturmian blocks}. The last block of a sequence $\{\mathcal{B}_k\}_{k=1}^n$, is said to be the Sturmian block associated to $\alpha$, and for them we will use the notation $\mathcal{B}(\alpha) = \mathcal{B}_n$. For numerical purposes, irrational numbers must be approximated by rationals approximants. Thus, if $\alpha$ is an irrational number, it can be approximated by the so--called $n$th convergent, say $\nu_n/\delta_n$,  where 
\begin{equation}
\frac{\nu_1}{\delta_1} =  \frac{1}{a_1} \ , \quad
\frac{\nu_2}{\delta_2} = \frac{1}{a_1 + \dfrac{1}{a_2}} \ , \
\ldots \ , \ \frac{\nu_n}{\delta_n} =  \frac{1}{a_1 + \dfrac{\phantom{1}}{\ddots  + \dfrac{1}{a_{n-1} + \dfrac{1}{a_n}}} }     \ .
\label{eq003}
\end{equation}
Numerators and denominators of the $n$th convergent can be determined separately throughout the schemes~\cite{Olds-1963}
\begin{eqnarray}
\nu_n &=& a_n \, \nu_{n-1} + \nu_{n-2} \ , \qquad \nu_{-1} =  1 \ , \quad \nu_{0} = 0   \ , \label{eq007a}  \\
\delta_n &=& a_n \, \delta_{n-1} + \delta_{n-2} \ , \qquad \delta_{-1} =  0 \ , \quad \delta_{0} = 1   \ . \label{eq007b}	
\end{eqnarray} 
 Let us denote by $\mathcal{N}_k$ to
  %
   the total number of symbols of the $k$th block, for $k \geq 0$. Due to the recursive relation of Eq.~\eqref{eq002},  at each step new $a_k$ blocks ot type $\mathcal{B}_{k-1}$ are added to the existing block $\mathcal{B}_{k-2}$, it follows immediately that
\begin{equation}
\mathcal{N}_{k} =  a_k \, \mathcal{N}_{k-1} + \mathcal{N}_{k-2}
\ , \quad 1 \leq k \leq n \ , \qquad
\mathcal{N}_{-1} = 1 \ , \quad
\mathcal{N}_{0} = 1  \ .\label{eq005} 
\end{equation}
It is straightforward that $\mathcal{N}_{k}  = \nu_k + \delta_k$, $k \geq -1$, something that will be used later. \\

  \begin{table}[ht]
	\centering
	\begin{tabular}{|c|c|ccccc|}
            \hline
            				&				$k$				&				1 			 & 				2				 & 					3					&				4	&		5			\\
\hline            				
$\alpha=3/11$   & $\mathcal{B}_{k}$		 &  $ppp \, q$			&		$pppq \, p$		  &		$pppqppppqp \, pppq$	& --- &		---			\\
							& $a_k$						  &				3				&				1				&				2						&		 ---	 &    ---			\\
							& $\nu_k/\delta_k$	  &				$1/3$		&			$1/4$			&			$3/11$					&		---		 &   ---			\\
							& $\mathcal{N}_k$		&			4				&				5				&				14						&	    --- 	 &   ---			\\
\hline            											
$\alpha=1/\phi$   & $\mathcal{B}_{k}$	 &  $pq$				 &		$pq \, p$		  &		$pqp \, pq$		&   $pqppq \, pqp$	&  
								$pqppqpqp	 \, pqppq$	\\
								& $a_k$						  &				1				&				1				&				1						&		 1		&		1				\\
								& $\nu_k/\delta_k$	&			$1/1$		&	$1/2$ 					& 			 $2/3$					& 		 $3/5$	&	   $5/8$ 				\\
								& $\mathcal{N}_k$	  &				1			&			3					&			5							&	8				&				13					\\
\hline            												
\end{tabular}
	\caption{Sturmian blocks, continued fractions and number of symbols associated to the numbers $\alpha = 3/11$ and $\alpha = 1/\phi$, where $\phi = (1+\sqrt{5})/2$ is the gold number. Both sequences start with $\mathcal{B}_{-1} = q$ and $\mathcal{B}_{0} = p$}
	\label{tab01}
\end{table}
Two illustrative examples are shown in Table~\ref{tab01} for the numbers $\alpha = 3/11 = [0;3,1,2]$ and $\alpha = 1/\phi = [0;1,1,\ldots]$, where $\phi = (1+\sqrt{5})/2$ is the gold number. This  latter corresponds to the well known Fibonacci sequence, although only the first five iterations have been listed. Note that for both numbers, the total amount of symbols is equal to $\mathcal{N}_n = \nu_n + \delta_n$, something that can be proved straightforward from Eqs.~\eqref{eq007a} and ~\eqref{eq007b}. Furthermore, it turns out that among the $\mathcal{N}_n$ symbols, there are $\nu_n$ are $q$'s and $\delta_n$ are $p$'s, being $\mathcal{B}_{-1} = q$ and $\mathcal{B}_{0} = p$.\\

We have defined Sturmian sequences for every real number between 0 and 1, including both limits. It is worthwhile to stop for a moment and describe what the sequences associated with these limits look like. First, $\alpha=0$ does not have strictly a continued fraction as shown in Eq.~\eqref{eq001} but can be considered the limit of $0 = \lim_{r\to \infty } [0;r]$ and therefore its associated sequence will also be the limit 
\begin{equation}
\mathcal{B}(\alpha=0) = \lim_{r\to \infty } \mathcal{B}(1/r) = \lim_{r\to \infty } p^r \, q = ppppp\ldots
\label{eq068}
\end{equation}
On the other side, the value $\alpha = 1$ has as (degenerated) continuous fraction $1 = 1/1$ and therefore its associated Sturmian sequence is
\begin{equation}
\mathcal{B}(\alpha=1) = pq \, pq \, pq \, pq \, \ldots
\label{eq069}
\end{equation}
As it will be seen later,  both limit values $\alpha = 0$ and $1$ are associated to very well-known systems: a
%
homogeneous medium and the periodic bi-layered structure, respectively. Now we have established how to construct Sturmian sequences from a simple two-symbol alphabet. In the next section we will describe how each Sturmian word or block can be associated with a mechanical structure.

\subsection{Quasiperiodic distribution of parameters}

\begin{figure}[ht]%
	\begin{center}
		\includegraphics[width=17cm]{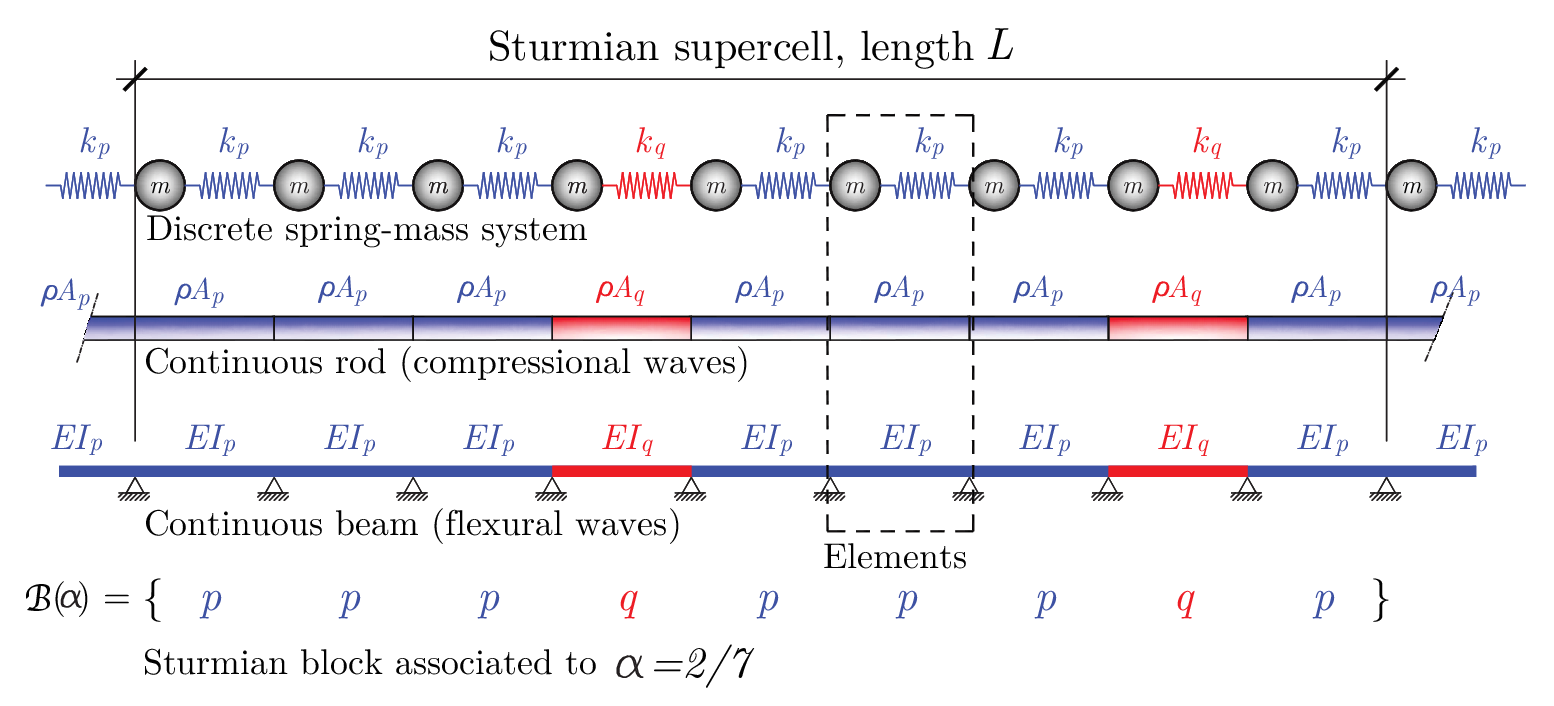} \\	
		\caption{Three different examples of dynamical systems based on Sturmian blocks for number $\alpha = 2/7 = [0;3,2]$. Above: a discrete spring-mass system, $\Theta \equiv k$ (spring coefficients). Middle: a continuous rod (axial waves), $\Theta \equiv \rho A$ (mass per unit of length). Bottom: a continuous beam (flexural waves), $\Theta \equiv EI$ (sectional bending stiffness) }%
		\label{fig03}%
	\end{center}
\end{figure}

Consider a dynamical system capable to propagate waves in one direction. Let us assume that system is formed by the concatenation of different elements, for instance masses, springs, rods, elastic supports, beams. All of these elements have mechanical, inertial and geometrical properties in the context of elastic waves. Thus, for instance, a discrete lumped mass system has as parameters the masses and the spring coefficients.  Given certain $\alpha \in \left[ 0,1\right] = [0;a_1,\ldots,a_n]$, then the Sturmian block associated to $\alpha$, $\mathcal{B}(\alpha) = \mathcal{B}_n$,  has exactly $N = \mathcal{N}_n$ symbols according to the pattern given by Eqs.~\eqref{eq002}  and~\eqref{eq005}. Our system arises as the periodic repetition of the block $\mathcal{B}(\alpha)$, which in turn is formed by $N$ elements. One of the parameters, say $\Theta$, can take only two values among the binary set $\{\theta_p,\theta_q\}$, meanwhile the rest of parameters remain constant from element to element along the block. Thus,  if $\Theta(j)$ denotes the value of the parameter of the $j$th element, with $1 \leq j \leq N$, then we have 
\begin{equation}
\Theta(j) =
\begin{cases}
\theta_p & \text{if the } j\text{th term of } \mathcal{B}(\alpha) \text{ is } p \\
\theta_q & \text{if the } j\text{th term of } \mathcal{B}(\alpha) \text{ is } q 
\end{cases}
\ , \quad 1 \leq j \leq N  \ .
\label{eq006}
\end{equation}
Since the system is formed by periodic repetition of the Sturmian block $\mathcal{B}(\alpha)$, then $\Theta(j+N) = \Theta(j)$ for $j > N$. In the Fig.~\ref{fig03} the building process of the system is illustrated with three examples, a discrete spring-mass system, a continuous rod (axial waves) and a continuous beam (flexural waves). The Sturmian pattern for the three systems is given by the number $\alpha = 2/7 = [0;2,3]$, resulting the block  $\mathcal{B}(\alpha) = pppqpppqp$. The elements of the discrete system are formed by one spring and one mass. The mass remains fixed but the rigidity of the spring assumes the roll of the parameter, i.e. $\Theta \equiv k$, and $k_p$ or $k_q$ are depending on the Sturmian sequence within $\mathcal{B}(\alpha)$. The second system (shown in the middle of Fig.~\ref{fig03}) represents a straight rod with density $\rho$, cross sectional area $A$ and Young modulus $E$. As known, the axial compressional waves propagate at a velocity $\sqrt{EA/\rho A}$. The infinite medium is structured into elements of length $l$. In the particular case of this example, the axial stiffness $EA$ is constant meanwhile the mass per unit of length $\rho A$ varies among two values $\{\rho A_p, \rho A_q\}$ as indicated in the the Sturmian block. The third example represents a beam on simple supports. Possible parameters which can be assigned to $\Theta$ are, for instance, $ \{\rho A, EI, GA_s \}$, where $EI$ and $GA_s$ are the sectional bending and shear stiffness, respectively. Even the span length between supports could be  changed from element to element obeying the Sturmian block pattern.  In Fig.~\ref{fig03} (bottom) the bending stiffness is assumed to take one of the two values $EI_p$ and $EI_q$ as prescribed in  $\mathcal{B}(\alpha)$. In this case, the three examples have $N=9$ elements which are repeated periodically. If $\alpha$ is an irrational number, theoretically the system is not periodic because $\mathcal{B}(\alpha)$ has infinite number of
symbols. In order to construct an achievable system, $\alpha$ has to be approximated by the $n$th convergent, i.e. $\alpha  \approx \nu_n / \delta_n$. As $n$ increases, the effects of quasiperiodicity become more relevant and visible in the system. One of the consequences is the selfsimilarity of the spectrum  as more terms of the sequence $\{a_n\}$ are added, something that can be visualized for the Fibonacci case in ref.~\cite{Gei2010}

Since from a practical point of view we can only form systems associated to rational numbers, the resulting parameter arrangement will result in a  so-called {\em supercell}  with $N = \mathcal{N}_n$ elements. This one will be repeated periodically in the same way as the associated Sturmian word, forming a mechanical waveguide.

\subsection{A geometrical interpretation}
\label{GeometricalInterpretation}

Consider a dynamical system where one of the parameter is tuned according to the two values $\Theta \in \{\theta_p,\theta_q\}$. As shown above, the system is an infinite media divided into elements. The Sturmian block $\mathcal{B}(\alpha)$ controls the pattern of assignation of $\Theta$ for each element. It is possible to relate the three numbers $\theta_p,\theta_q,\alpha$ with a geometrical construction, providing a graphical interpretation of the system and in turn with some analytical implications. The following developments make sense when $\alpha$ is a rational number. Otherwise, it has already been seen that the system cannot be evaluated computationally, except if it is considered as the limit of a sequence of finite systems associated with the different convergents. \\
\begin{figure}[h]
	\begin{center}
		\includegraphics[width=13cm]{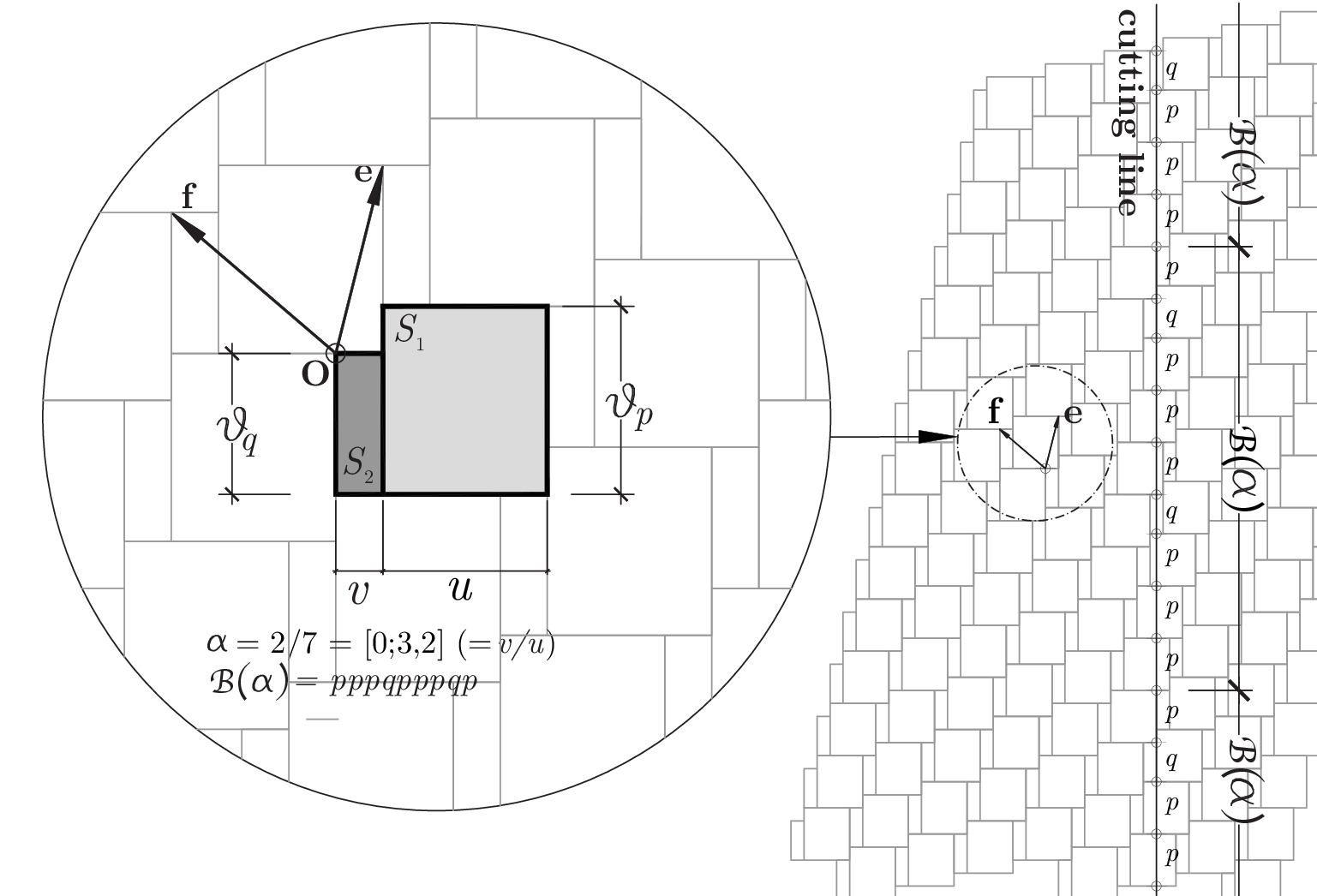}
		\caption{Two rectangles $S_1 (u \times \theta_p)$ and $S_2(v \times \theta_q)$ forming a tiling by periodic translation of basis  $\mathbf{e}=(v,\theta_p), \ \mathbf{f}=(-u,\theta_q)$. The examples shown has been built for $\alpha = v/u = 2/7 = [0;3,2]$. }
	\label{fig01}
	\end{center}
\end{figure}

\begin{figure}[ht]%
	\begin{center}
		\includegraphics[width=13cm]{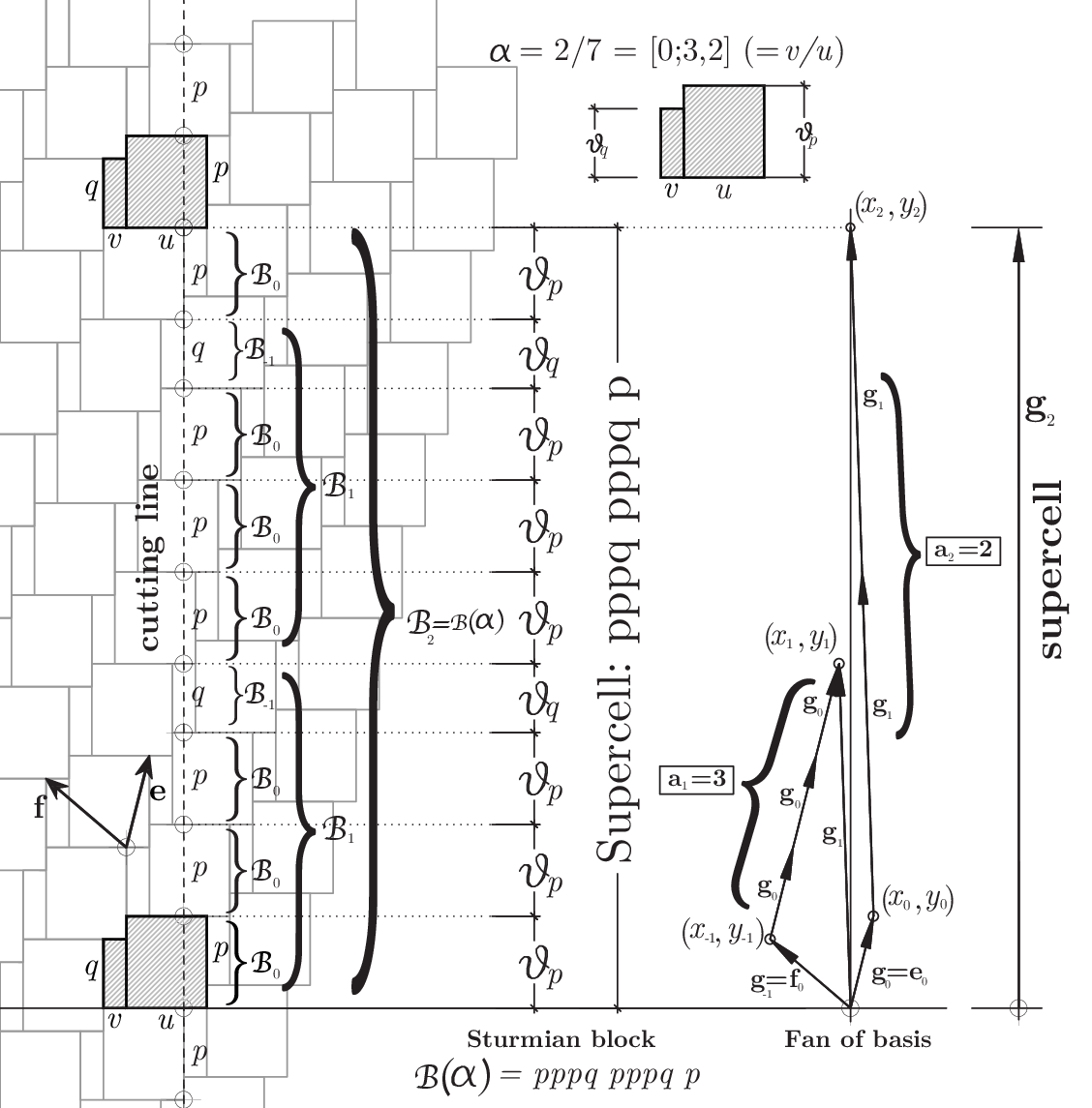} \\	
		\caption{Lattice plane tiling using rectangles $\mathcal{S}_1$ and $\mathcal{S}_2$. Sturmian blocks associated to $\alpha = 2/7 = [0;3,2]$ and the corresponding Sturmian-like spring-mass chain. Right: sequence of bases $\{\mathbf{g}_k\}_{k=1}^n$ and {\em length} of the supercell in terms of the quasiperiodic parameter $\sum_{j=1}^n\Theta(j) = \nu_n \theta_q + \delta_n \theta_p$}%
		\label{fig02}%
	\end{center}
\end{figure}

Assume that $\alpha = [0;a_1,\ldots,a_n] \in [0,1]$ can be written as the quotient of two coprime integers $\alpha = v/u$, with $v \leq u$. Let us consider two rectangles $S_1, S_2 \subset \mathbb{R}^2$ with widths $u$ and $v$ and heights  $\theta_p$ and $\theta_q$, respectively (see Fig \ref{fig01}). We can now form a basis of vectors $\{\mathbf{e},\mathbf{f} \} $ where
\begin{equation}
\mathbf{e} =  (v,\theta_p) \ , \qquad \mathbf{f}=(-u,\theta_q)  \ .
\label{eq012}
\end{equation}
Translating repeatedly the figure $S_1 \cup S_2$ in the both directions given by the vectors $\mathbf{e}$ and $\mathbf{f}$, a bidimensional tiling arises. We are interested in the pattern of the cutting points of any vertical line with horizontal sides of the tiling. It is clear that the distance between two consecutive intersecting points is either $\theta_p$ or $\theta_q$. It turns out that the sequence of the resulting segments, for instance $\theta_p \, \theta_p \, \theta_p \, \theta_q \ldots$ coincides with that one given by the Sturmian block. The fact that $\alpha = v/u$ is a rational number makes the pattern to periodically repeat, just as our unit cell is replicated in the infinite media. This property can be proved with help of the following sequence of vectors  $\{\mathbf{g}_k\}$ defined recursively from the basis $\{\mathbf{e,f}\}$ and from the continued fraction of $\alpha$.
\begin{equation}
\mathbf{g}_{k} = a_k \, \mathbf{g}_{k-1} + \mathbf{g}_{k-2} 
\ , \quad  0 \leq k \leq n
\ , \qquad 
\mathbf{g}_{-1} = \mathbf{f} \ , \quad 
\mathbf{g}_{0} = \mathbf{e}  \ .
\label{eq011}
\end{equation}
It can be demonstrated straightforward by induction that $\mathbf{g}_k = \nu_k \, \mathbf{f}  + \delta_k \, \mathbf{e}$ since both $\nu_k$ and $\delta_k$ also can be formed under the same pattern, see Eqs.~\eqref{eq007a} and~\eqref{eq007b}. Using the components of $\mathbf{e} $ and $\mathbf{f}$ from Eq. \eqref{eq012} and denoting by $(x_k,y_k)=\mathbf{g}_k$, it yields
\begin{equation}
x_k = \delta_k \, v - \nu_k \, u \quad , \quad 
y_k = \delta_k \, \theta_p + \nu_k \,\theta_q  \quad , \quad  0 \leq k \leq n  \ .
\label{eq013}
\end{equation}
Thus, in the $k$th step, the block $\mathcal{B}_k$ has $\mathcal{N}_k = \nu_k + \delta_k$ symbols among which $\nu_k$ are $q$'s and $\delta_k$ are $p$'s. In particular, for the last step,  if $\alpha = \nu_n/\delta_n$ (rational number), then $x_n = \delta_n \, u \left( \alpha - \nu_n/\delta_n\right) =0$. In other case, $\alpha \approx \nu_n / \delta_n$ represents an approximant and therefore $x_n \neq 0$. Moreover, the vertical component $y_n$ represents graphically the segment length of the window $y_n = \delta_n (\theta_p + \alpha \theta_q)$ in the tiling (see Fig. \ref{fig02} ). In terms of the parameters of the system, we have $y_n = \sum_{j=1}^N \Theta(j)$, where $N$ denotes the size of the sequence. Since $N = \nu_n + \delta_n = \delta_n (1 + \alpha)$ then the following relationship between the parameters of the system and the number $\alpha$ can be established
\begin{equation}
\sum_{j=1}^N \Theta(j) = N \frac{\theta_p + \alpha \, \theta_q}{1 + \alpha}  \ .
\end{equation}
In the Fig.~\ref{fig02} the geometrical interpretation of the Sturmian block is illustrated for the number $\alpha = 2/7 = [0;3,2]$. Since the continued fraction has two terms ($n=2$), the last vector $\mathbf{g}_2$ results to be purely vertical with length equal to $2\theta_q + 7 \theta_p$. The quasiperiodic distribution of the parameter $\Theta$ can be visualized following the cutting points of any vertical line with the horizontal sides of the rectangles $S_1,S_2$. Similarly the different vectors of the sequence $\{\mathbf{g}_k\}$ form a fan which tends to be vertical. Furthemore,  if $\alpha$ is an irrational number, then the greater $n$, the closer to the vertical line is $\mathbf{g}_n$. In the latter case, the dynamical properties of the system must be approximated by the $n$th convergent. \\

More details about the relationship between Sturmian words and tilings formed the so-called bi-partition of the torus can be found in the references \cite{Siemaszko2011,Siemaszko2012}. There exist other geometrical approaches to obtain quasiperiodic patterns, for instance the cutting sequences~\cite{Series-1985a,Crisp-1993}, the cut and project scheme~\cite{Bellissard-1989,Godreche-1993} or the rotation-projection algorithm~\cite{Apigo-2018,Ruzzene-2019}.


\section{Spectrum of Sturmian structured media}

One of the most relevant consequences of wave propagation in structured media is the emergence of bandgaps in the frequency spectrum. In periodic media, a proper design of the unit cell can result in optimized location of bandgaps or passbands for practical interest. On other side, quasiperiodic media, like for example Fibonacci sequence--based systems, exhibit self-similarity of the spectrum ~\cite{Macia-2009,Velasco-2001,Kolar-1992}.  In this article, we propose to study one-dimensional quasiperiodic systems formed by structural elements. The proposed method allows to associate a generating parameter $\alpha \in \left[0,1\right]$ with each system. By sweeping out the values of such generator, we can form a family of structures with tailored properties as for instance can be the dispersion relations or resonances.
 We seek to relate this generating parameter $\alpha$ to the  admitted frequencies in the system by means of the so--called {\em bulk spectrum}.
%

\subsection{Dispersion relation and bulk spectrum}

\begin{figure}[ht]%
	\begin{center}
		\includegraphics[width=16cm]{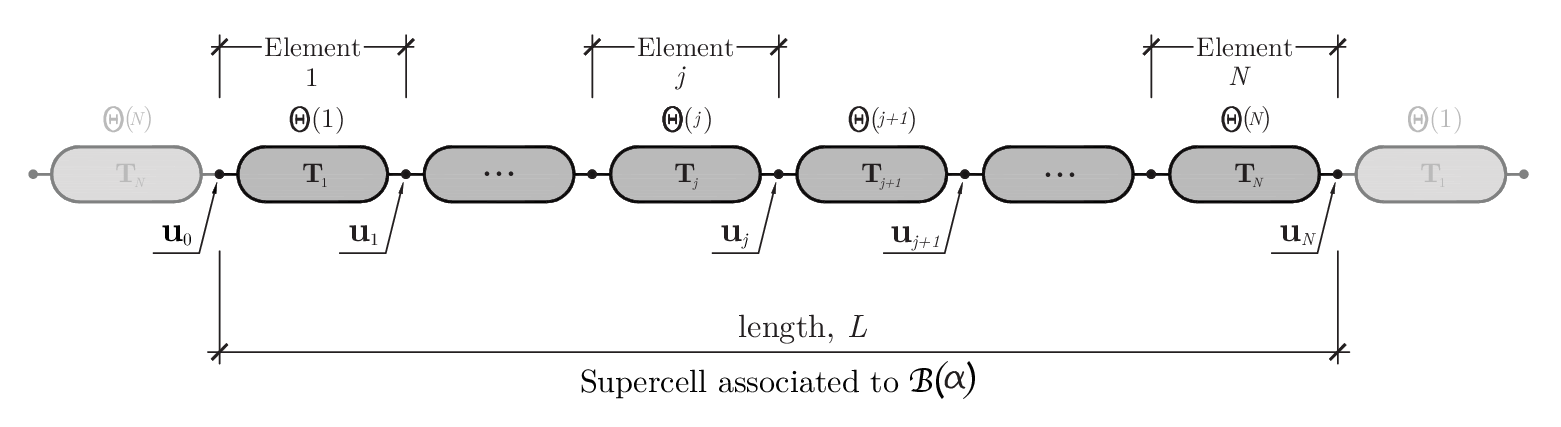} \\	
		\caption{A one-dimensional Sturmian structured system associated to the number $\alpha$. The binary parameter $\Theta(j) \in \{\theta_p,\theta_q\}$ changes its value according to the Sturmian pattern given by the block $\mathcal{B}(\alpha)$}%
		\label{fig04}%
	\end{center}
\end{figure}

Dispersion relation relates  frequencies and wavenumbers that  
can be exited in an infinite medium. For one--dimensional propagation, the transfer matrix method (TMM) results suitable to describe the transmission of the system variables from one element to the next one and ultimately between supercells. Energy transport mechanisms, dispersion curves, density of states or transmission-reflexion coefficients can be addressed using this approach. The TMM allows to express the state variables of the problem associated to a point of the system from those of another point by means of product of matrices, collecting the system properties between both points. Denote by $\bm{\mathfrak{u}}(x,t)$ the state vector in time-domain at position $x$ and at instant $t$. In general, this vector contains both node displacements and internal forces of the system. Considering harmonic motion, we can write $\bm{\mathfrak{u}}(x,t) = \mathbf{u}(x) \, e^{\text{i}\omega t}$. Let us consider two points in the system $x_j$ and $x_k$ and denote $\mathbf{u}_j = \mathbf{u}(x_j) , \ \mathbf{u}_k = \mathbf{u}(x_k)$. Then, the transfer matrix of the system between nodes $j$ and $k$, such that $x_j < x_k$ is a square matrix $\mathbf{T}_{jk}$ such that 
\begin{equation}
\mathbf{u}_{k} = \mathbf{T}_{jk} \, \mathbf{u}_{j}  \ . 
\label{eq014} 
\end{equation}
Some relevant properties of the transfer matrix~\cite{Rui-2019} which will be used later:
\begin{itemize}
	\item It is unimodular, i.e. $\det \left( \mathbf{T}_{jk} \right) = 1$.
	\item It depends on the elastodynamical properties of the system within $x_j \leq x \leq x_k$ and on the frequency $\omega$. 
	\item If $x_1 < x_2 < \cdots < x_{j-1} < x_j$ then $\mathbf{T}_{1j} =  \mathbf{T}_{j-1,j}  \, \mathbf{T}_{j-2,j-1} \cdots \mathbf{T}_{12}$.
\end{itemize}
As described above, Sturmian structured systems are constructed on the assumption that certain element parameter $\Theta$ is tunned following the Sturmian pattern $\mathcal{B}(\alpha)$ associated to certain number $\alpha = [0;a_1,\ldots,a_n]$. In the Fig.~\ref{fig04} a schematic sketch of a Sturmian system is shown, where the supercell given by $\mathcal{B}(\alpha)$ is periodically repeated. Denoting by $\mathbf{T}_j$ to the transfer matrix between nodes $j-1$ and $j$, then we can write the relationship between the state vectors at the two ends of the unit cell as

\begin{equation}
\mathbf{u}_N = \left( \mathbf{T}_N \cdot \cdots \cdot \mathbf{T}_1 \right) \, \mathbf{u}_0 \equiv \bm{\mathcal{T}}(\alpha) \,  \mathbf{u}_0 \ .
\label{eq017}
\end{equation}

The above expression holds for any one-dimensional dynamic model, regardless of the algorithm used for its construction. In the particular case of Sturmian structured systems, we have already seen above that the block $\mathcal{B}(\alpha)$ set the value of the property $\Theta$. Since each block emerges from concatenation of previous blocks according to the rule \eqref{eq002}, then the transfer matrix associated to each block can also be determined in a recursive way in terms of matrix product.  Indeed, denoting by  $\bm{\mathcal{T}}_k$ to the transfer matrix associated to the $k$th Sturmian block $\mathcal{B}_k$, then the way in which they can recursively be determined is shown in Table~\ref{tab03}. 

\begin{table}[ht]
	\centering
	\begin{tabular}{|r|rl|rl|}
		\hline
		Order, $k$		&		&Sturmian Block, $\mathcal{B}_k$		&		& Transfer Matrix, $\bm{\mathcal{T}}_k$		\\
		\hline
		$-1$		 &		 $\mathcal{B}_{-1} = $ & $q$						 &		 $\bm{\mathcal{T}}_{-1} =$ &$\mathbf{T}_q$		 \\		 
		$0$		 &		 $\mathcal{B}_{0}=$ & $p$						 &		 $\bm{\mathcal{T}}_{0} = $ & $\mathbf{T}_p$		 \\
		$1$      &       $\mathcal{B}_{1} = $ & $\mathcal{B}_{0}\stackrel{a _1}{\ldots}  \mathcal{B}_{0} \, \mathcal{B}_{-1} \,  $	   
		 				 &	$\bm{\mathcal{T}}_{1} =$ &$ \bm{\mathcal{T}}_{-1} \, \bm{\mathcal{T}}_{0}^{a_1} $		 \\		     
		 $\ldots$ &   & $   \ldots $ 					  &	 $\ldots $ &		 \\		     
		 $n$		  &       $\mathcal{B}_{n} =$ & $\mathcal{B}_{n-1}\stackrel{a _n}{\ldots}  \mathcal{B}_{n-1}  \, \mathcal{B}_{n-2}  $	 
		 				 &	$\bm{\mathcal{T}}_{n} =$& $ \bm{\mathcal{T}}_{n-2} \, \bm{\mathcal{T}}_{n-1}^{a_n} $	\\	 		     	
		 \hline	 		 
\end{tabular}
	\caption{Transfer matrix of the $k$th Sturmian block associated to $\alpha = [0;a_1,a_2,\ldots,a_n]$. Matrices $\mathbf{T}_p$ and $\mathbf{T}_q$ denote respectively the transfer matrices of elements with parameters $\theta_p$ and $\theta_q$}
	\label{tab03}
\end{table}

\begin{eqnarray}
\bm{\mathcal{T}}_k &=&  \bm{\mathcal{T}}_{k-2}   \bm{\mathcal{T}}_{k-1}^{a_k}  \ ,  \quad 1 \leq k \leq n  \ ,  \nonumber \\
\bm{\mathcal{T}}_{-1} &=&  \mathbf{T}_q \ , \quad \bm{\mathcal{T}}_{0} =  \mathbf{T}_p  \ .
\label{eq019}
\end{eqnarray}
The TM of the unit cell is then $\bm{\mathcal{T}}(\alpha) = \bm{\mathcal{T}}_n$ and relates the state variables $\mathbf{u}_0$ and $\mathbf{u}_N$ yielding
\begin{equation}
\mathbf{u}_N = \bm{\mathcal{T}}(\alpha)  \, \mathbf{u}_0  \ .
\label{eq020}
\end{equation}
Applying the Bloch theorem to the unit cell we know that that $\mathbf{u}_N = e^{\text{i}  \kappa L} \mathbf{u}_0$, thus Eq~\eqref{eq020} can be written  then as the linear eigenvalue problem
\begin{equation}
\left[ \bm{\mathcal{T}}(\alpha)  - \lambda \, \mathbf{I} \right] \, \mathbf{u}_0 = \mathbf{0}  \ , 
\label{eq021}
\end{equation}
where the parameter is $\lambda = e^{\text{i}  \kappa L}$. As known, the TM depends on the frequency $\omega$. Certain wave of frequency $\omega$ is said to be admitted in the medium if there exist real solution for the wavenumber $\kappa$ from Eq.~\eqref{eq021}. In such a case, the frequency lies within a passband. Othewise, it is in a bandgap or stopband, where the wave cannot be transmitted in the medium because is evanescent with an exponentially decaying amplitude (complex wavenumber).  Passbands and stopbands can be visualized in the dispersion curves which arise as solutions of the equation $\det \left[  \bm{\mathcal{T}}(\alpha)  - \lambda \, \mathbf{I}\right]=0$. In most structural models of rods and beams the transfer matrices are $2 \times 2$ or $4 \times 4$ in size. For them, closed forms for the dispersion relations can be derived. \\

If $\bm{\mathcal{T}}(\alpha) $ is a $2 \times 2$ matrix then the characteristic polynomial of Eq.~\eqref{eq021} is
\begin{equation}
\det \left[  \bm{\mathcal{T}}(\alpha)  - \lambda \, \mathbf{I}\right] =
\lambda^2 - \tr \left[ \bm{\mathcal{T}}(\alpha) \right] \, \lambda + \det \left[ \bm{\mathcal{T}}(\alpha)\right]   \ , 
\label{eq023}
\end{equation}
where $\tr(\bullet)$ stands for the matrix trace operator. The fact that transfer matrix are unimodular notably simplifies the problem resulting, after some straight operations, the final expression for $2\times 2$ TM spectrum
\begin{equation}
\cos \left(  \kappa L \right) = \frac{1}{2} \tr \left[ \bm{\mathcal{T}}(\alpha)  \right]   \ .
\label{eq022}
\end{equation}
This approach can be extended to consider system  described under $4 \times 4 $ transfer matrices~\cite{Macia-2009}. Indeed, in such case the characteristic polynomial results
\begin{equation}
\det \left[  \bm{\mathcal{T}}(\alpha)  - \lambda \, \mathbf{I}\right] =
\lambda^4 - c_1  \lambda^3 + c_2 \lambda^2 - c_3 \lambda + c_4 = 0  \ , 
\label{eq024}
\end{equation}
where
\begin{eqnarray}
c_1 &=& \tr \left[ \bm{\mathcal{T}}(\alpha) \right]  \ , \nonumber \\
c_2 &=& \frac{1}{2} \left(  \tr^2 \left[ \bm{\mathcal{T}}(\alpha) \right] - \tr \left[ \bm{\mathcal{T}}^2(\alpha) \right] \right)  \ , \nonumber \\
c_3 &=& \frac{1}{6} \left(  \tr^3 \left[ \bm{\mathcal{T}}(\alpha) \right] - 
										3	\tr \left[ \bm{\mathcal{T}}^2(\alpha) \right] \, 
											\tr \left[ \bm{\mathcal{T}}(\alpha) \right] +
										2	\tr \left[ \bm{\mathcal{T}}^3(\alpha) \right] \right)  \ , \nonumber \\
c_4 &=& \det \left[ \bm{\mathcal{T}}(\alpha)\right]										 \ .
\label{eq025}
\end{eqnarray}
If $\kappa$ is the wavenumber of a solution representing a Bloch wave traveling towards the right, then $\lambda^{-1} = e^{-\text{i} \kappa L}$ must be also a solution traveling leftwards. Therefore, Eq.~\eqref{eq024} holds also for $\lambda^{-1}$, yielding
\begin{equation}
\det \left[  \bm{\mathcal{T}}(\alpha)  - \lambda \, \mathbf{I}\right] =
\lambda^{-4} - \tr \left[ \bm{\mathcal{T}}(\alpha) \right]  \lambda^{-3} + c_2 \lambda^{-2} - c_3 \lambda^{-1} + c_4 = 0  \ .
\label{eq026}
\end{equation}
Multiplying by $\lambda^4$ and identifying coefficients with Eq.~\eqref{eq024} one gets $c_3 = c_1$ and $c_4 = 1$. Moreover, using these results and dividing Eq.~\eqref{eq024} $\lambda^2$, the quartic equation can be reduced to the quadratic polynomial
\begin{equation}
s^2 - \tr \left[ \bm{\mathcal{T}}(\alpha) \right] \, s + c_2 -2 = 0  \ , 
\label{eq027}
\end{equation}
where 
\begin{equation}
s = \lambda + \lambda^{-1} = e^{\text{i} \kappa L} + e^{-\text{i} \kappa L} = 2 \cos (\kappa L) = 
\frac{\tr \left[ \bm{\mathcal{T}}(\alpha) \right] \pm \sqrt{\tr^2\left[ \bm{\mathcal{T}}(\alpha) \right] - 4 (c_2-2)}    }{2}  \ .
\label{eq028}
\end{equation}
Finally, after some simplifications the spectrum yields
\begin{equation}
\cos (\kappa L) = \frac{1}{4} \left[ \tr \left[ \bm{\mathcal{T}}(\alpha) \right] \pm 
													\sqrt{2 \tr \left[ \bm{\mathcal{T}}^2(\alpha) \right] 
													- \tr^2\left[ \bm{\mathcal{T}}(\alpha) \right] + 8 }     \right]  \ .
\label{eq029}												
\end{equation}
The two solutions obtained lead to two dispersion branches related to waves of different nature in the model. Thus, for instance, in the case of Timoshenko beams, (revisited below as an example of $4\times 4$ TM), both solutions correspond to the spectrum of pure bending and shear waves associated with each frequency. \\

From both Eqs~\eqref{eq022} and~\eqref{eq029} the wavenumber $\kappa(\omega)$ can be expressed analytically as function of frequency. Admitted frequencies are those values of $\omega$ which lead to a real wavenumber $\kappa(\omega)$. For $2 \times 2$ TM, this can be reduced to the condition $-2 \leq \tr \left[ \bm{\mathcal{T}}(\alpha) \right] \leq 2$. For $4 \times 4$ TM bandgaps are defined as those frequencies which make the right hand side of Eq. ~\eqref{eq029} to be higher than 1 in absolute value. Collecting the admitted frequencies, they can be arranged along a line so that  passbands are depicted as segments and stopbands are the bandgaps between them. Repeating the process for the whole range of $\alpha$ the passbands and stopbands forme a figure, called {\em bulk spectrum} (BS). The finer the discretization of the interval $[0,1]$, the better resolution of this graphical figure, which allows to visualize at a single glance the quasiperiodic profile of our system based on the Sturmian sequences. In the following section we will present some properties of BS that can be established a priori and that will be tested later in the numerical examples.


\subsection{Spectrum properties and selfsimilarity}
\label{selfsimilarity}

Given certain $\alpha = \nu_n / \delta_n$ in the range $0 \leq \alpha \leq 1$, the mechanical system constructed from the Sturmian Block has exactly $N(\alpha) = \nu_n + \delta_n$ elements, number that coincides with the number of symbols in the word $\mathcal{B}(\alpha)$. 
Thus, for instance,  systems associated to simple fractions like for instance $\alpha = 1/2, \ 1/3, \ 3/4$ will present 3, 4 and 7 elements respectively. The number of passbands/stopbands and their distribution depends directly on the arrangement of elements in the supercell. However, for an irrational number $\alpha$, the block $\mathcal{B}(\alpha)$  results in an infinite word and therefore it will present infinite passbands and bandgaps. For certain type of irrationals, the frequency spectrum results to be selfsimilar~\cite{Macia-2009}. Since both type of spectrums, associated to rational and irrational numbers, coexist in the same bulk spectrum, our goal is trying to explain the geometrical structure and to justify the selfsimilarity properties observed in the numerical examples. \\

Consider two points $a$ and $b$ inside the interval $[0,1]$ defined as 
\begin{equation}
a = \left[ 0;a_1,\ldots,a_n \right] = 
 \frac{1}{a_1 + \dfrac{1}{\ddots  + \dfrac{1}{a_{n-1} + \dfrac{1}{a_n}}} } = \frac{\nu_n}{\delta_n}   \  , \quad 
b = \left[ 0;a_1,\ldots,a_{n-1}\right] =
\frac{1}{a_1 + \dfrac{1}{\ddots  + \dfrac{1}{a_{n-1} }}}  = \frac{\nu_{n-1}}{\delta_{n-1}}   \ .
\label{eq031}
\end{equation}
The matrices $\bm{\mathcal{T}}(a)$ and $\bm{\mathcal{T}}(b)$ stand for the TM's of the Sturmian systems associated to both numbers. We know that both matrices are determined as the last term of the  recursion formula~\eqref{eq019}, namely
\begin{eqnarray}
\bm{\mathcal{T}}(a) &=& \bm{\mathcal{T}}_{n} =  \bm{\mathcal{T}}_{n-2} \,    \bm{\mathcal{T}}_{n-1}^{a_n}   \ , \nonumber   \\
\bm{\mathcal{T}}(b) &=& \bm{\mathcal{T}}_{n-1} =  \bm{\mathcal{T}}_{n-3}   \bm{\mathcal{T}}_{n-2}^{a_{n-1}}  \ .
\label{eq032}
\end{eqnarray}
Although there are infinitely many numbers between $a$ and $b$ and therefore infinitely many systems, there is a family of systems that can be represented mathematically by a numerable set. This family allows us to deduce some interesting properties that help to predict the behavior of the spectrum. Let $\{\alpha_r\}_{r \geq 0}$ be a sequence of real numbers defined as
\begin{equation}
\alpha_r = \left[  0;a_1,\ldots,a_n + r \right] = 
 \frac{1}{a_1 + \dfrac{1}{\ddots  + \dfrac{1}{a_{n-1} + \dfrac{1}{a_n + r}}} }   \ .
 \label{eq052}
\end{equation}
It is straightforward that
\begin{equation}
\lim_{r \to 0}  \alpha_r =  a \quad , \quad  \lim_{r \to \infty} \alpha_r  =  b 
\label{eq053}
\end{equation}
In addition, $\alpha_r$ lies between $a$ and $b$. Indeed, since $a = \nu_n/\delta_n$ and $b = \nu_{n-1} / \delta_{n-1}$, then by Eqs.~\eqref{eq007a} and~\eqref{eq007b} we have $\alpha_r = \nu(r) / \delta(r)$ with
\begin{eqnarray}
\nu(r) &=& (a_n + r) \, \nu_{n-1} + \nu_{n-2} = \nu_n + r \, \nu_{n-1}  \ ,  \nonumber \\
\delta(r) &=&  (a_n + r) \, \delta_{n-1} + \delta_{n-2} = \delta_n + r \, \delta_{n-1} \label{eq033}  \ .
\end{eqnarray}
Hence,
\begin{equation}
\alpha_r = \frac{\nu(r)}{\delta(r)} = \frac{\nu_n + r \, \nu_{n-1} }{\delta_n + r \, \delta_{n-1}} 
			= \frac{a \, \delta_n + r \, b \, \delta_{n-1} }{\delta_n + r \, \delta_{n-1}} = \zeta_r \, a + (1 - \zeta_r) \, b  \ , 
			\label{eq034}
\end{equation}
where $\zeta_r = \delta_{n} / (\delta_{n} + r \delta_{n-1})$, with $0 < \zeta_r \leq 1$. Eqs.~\eqref{eq034} shows clearly that $\alpha_r$ lies between the two points no matter the order, because a priori it is not know whether $a < b$ or $b<a$. \\

Let us denote by $\bm{\tau}_r = \bm{\mathcal{T}}(\alpha_r)$ to the transfer matrix of the Sturmian system associated to $\alpha_r$. By definition of $ \bm{\mathcal{T}}(\alpha_r)$ it yields
\begin{equation}
\bm{\tau}_r = \bm{\mathcal{T}}_{n-2} \,    \bm{\mathcal{T}}_{n-1}^{a_n+r} 
			     =  \left(\bm{\mathcal{T}}_{n-2} \,    \bm{\mathcal{T}}_{n-1}^{a_n} \right)\,  \bm{\mathcal{T}}_{n-1}^{r} 
			     =  \bm{\mathcal{T}}(a) \, \bm{\mathcal{T}}^r(b)  \ .
			     \label{eq036}
\end{equation}
This expression exhibits the nature of the system associated with $\alpha_r$ as the concatenation of the systems $\mathcal{B}(a)$ and $\mathcal{B}(b)$, this latter $r$ times. 
 \begin{equation}
 \mathcal{B}(\alpha_r) = \mathcal{B}(b) \stackrel{r}{\ldots}  \mathcal{B}(b)  \, \mathcal{B}(a)  \ .
 \label{eq035}
 \end{equation}
 The effect of introducing the system $\mathcal{B}(a)$ into a system of the form $ \mathcal{B}(b) \stackrel{r}{\ldots}  \mathcal{B}(b)  $ with $r \gg 1$ is to open the passbands of the latter with small bandgaps associated to the former. The closer we are to $b$ (the greater $r$)  the more bandgaps will be open but also the narrower. The interesting point about this behavior is that it occurs at all scales independently of the two points $a$ and $b$ chosen, as long as they are the initial and limit numbers  of the sequence $\{\alpha_r\}$. We are therefore faced with a fractal pattern in the figure resulting from the spectrum. Mathematically, this relationship between the number of passbands and the distance between $\alpha_r$ and $b$ can be visualized in a simple expression. Indeed, consider the parameter $\zeta$ introduced in Eq.~\eqref{eq034} as the relative distance between $\alpha_r$ and $b$. In fact, it yields
 \begin{equation}
 \zeta_r = \frac{\alpha_r - b}{a - b} = \frac{\delta_n}{\delta_{n} + r \delta_{n-1}}  \ .
 \label{eq037}
 \end{equation}
Now, the number of branches in the dispersion relation of the system associated to $\alpha = [0;a_1,\ldots,a_n]$ is directly proportional to the amount of distinct elements  $N(\alpha) = \nu_n + \delta_n$. Thus, for the numbers $a$, $b$ and $\alpha_r$, we have 
\begin{eqnarray}
N(a) &=& \nu_n + \delta_n = (1 + a) \, \delta_n   \ ,  \nonumber \\
N(b) &=& \nu_{n-1} + \delta_{n-1} = (1 + b) \, \delta_{n-1}   \ ,  \nonumber \\
N(\alpha_r) &=& \nu(r) + \delta(r) = (1 + \alpha_r) \, \delta(r) = N(a) + r N(b)   \ , \label{eq038}
\end{eqnarray}
where the latter expression holds directly from Eqs.~\eqref{eq033}. Plugging Eq.~\eqref{eq038} into Eq.~\eqref{eq037} and after some straight operations we have finally
\begin{equation}
\zeta_r = \frac{\alpha_r - b}{a - b} =  \frac{1}{1 + \displaystyle \frac{1+a}{1+b} \, \frac{N(\alpha_r)-N(a)}{N(a)}}  \ .
\label{eq039}
\end{equation}
Moreover, when $a$ and $b$ lies very closed each other it is verified that
\begin{equation}
\zeta_r = \frac{\alpha_r - b}{a - b} \approx  \frac{N(a)}{N(\alpha_r)} \quad  \text{if} \ \left| a - b \right| \ll 1  \ .
\label{eq040}
\end{equation}
The system $\mathcal{B}^r(b)$, formed as periodic repetition of $\mathcal{B}(b)$  $r$ times, has a larger number of branches in the dispersion relation with
respect to that of  $\mathcal{B}(b)$, but does not add new bandgaps in the spectrum of admitted frequencies, these are simply new foldings of the Brillouin region. However, $\mathcal{B}^r(\alpha_r) = \mathcal{B}^r(b) \mathcal{B}(a)$ results from embedding $\mathcal{B}(a)$ within $\mathcal{B}^r(b)$, and the new heterogeneity induces new bandgaps~\cite{Macia-2009} in the branches of $\mathcal{B}^r(b)$. Eq.~\eqref{eq040}, relates the distance $\left| \alpha_r - b\right| $ and the ratio between the size of the systems, which in turn is somehow equivalent to the proportion of passbands of $\mathcal{B}(a)$ and $\mathcal{B}(\alpha_r)$. This relationship between the geometric distance on the axis of the alpha parameter and the structure of the spectrum will become clear in the numerical examples.  Let us see that we can analytically extract the spectrum as an explicit function of $r$ for the particular case of $2 \times 2$ transfer matrices. \\

Given $0 < a \leq 1$, then the spectrum of admitted waves in the system $\mathcal{B}(\alpha_r)$ is given by those values of the frequency $\omega$ which lead to real eigenvalues of the wavenumber $\kappa(\omega)$ in the eigenproblem
\begin{equation}
\left[ \bm{\tau}_r(\omega)  - \lambda \, \mathbf{I} \right] \, \mathbf{u}_0 = \mathbf{0}  \ , 
\label{eq041}
\end{equation}
where $\lambda = e^{\text{i} \kappa L}$ and $\bm{\tau}_r(\omega) = \bm{\mathcal{T}}(a) \, \bm{\mathcal{T}}^r(b)$ is the TM given by Eq.~\eqref{eq036}, highlighting the frequency dependency.  Some properties of the $2 \times 2$--size unimodular matrices allow us to determine a closed form of the dispersion relation associated to each number of the sequence. As stated above, the dispersion relation of a system modeled with $2 \times 2$ matrices is closely related to the trace of the TM, so that the spectrum of $\alpha_r$ can be represented by the set
\begin{equation}
\{\omega \in \mathbb{R}: -1 \leq \tr \left[ \bm{\tau}_r(\omega) \right] /2 \leq 1 \}    \ .
\label{eq044}
\end{equation}

From Eq.~\eqref{eq036} it yields
\begin{eqnarray}
\bm{\tau}_{r+1} &=& \bm{\mathcal{T}}(a) \, \bm{\mathcal{T}}^{r+1}(b) = \bm{\tau}_{r} \,  \bm{\mathcal{T}}(b)  \ ,  \label{eq043a}  \\
\bm{\tau}_{r-1} &=& \bm{\mathcal{T}}(a) \, \bm{\mathcal{T}}^{r-1}(b) = \bm{\tau}_{r} \,  \bm{\mathcal{T}}^{-1}(b)  \ . \label{eq043b}
\end{eqnarray}
For the following developments we need two properties of the 2nd order unimodular square matrices enabling to derive closed expressions for the spectrum associated to $\alpha_r$. Consider $\mathbf{A}, \mathbf{B} \in \mathbb{R}^{2 \times 2}$ and $\det \mathbf{A} = \det \mathbf{B} =  1$ then 
\begin{itemize}
	\item [(i)] $\tr \left[\mathbf{A} \mathbf{B}\right] = \tr \left[\mathbf{A} \right] \, \tr \left[ \mathbf{B}\right] - \tr \left[\mathbf{A} \mathbf{B}^{-1}\right]$. 
	\item [(ii)] $\mathbf{A}^{r+1} = \mathcal{U}_{r}(z) \ \mathbf{A} -  \mathcal{U}_{r-1}(z) \ \mathbf{I}_2 \ , \quad z = \tr \mathbf{A}/2 \ , \ r = 1,2,3,\ldots$
\end{itemize}
where $ \mathcal{U}_{r}(z)$ the $r$th order Chevyshev polynomial of second kind. In Eqs.~\eqref{eq043a} and~\eqref{eq043b} the trace of $\bm{\tau}_r$ results to be the trace of a product of unimodular matrices. Thus, taking traces on Eq.~\eqref{eq043a} and making use of Eq.~\eqref{eq043b}
\begin{equation}
\tr \left[  \bm{\tau}_{r+1}   \right] = \tr \left[  \bm{\tau}_{r} \,  \bm{\mathcal{T}}(b)  \right] = 
 \tr \left[  \bm{\tau}_{r} \right] \, \tr \left[  \bm{\mathcal{T}}(b)  \right]  - \tr \left[  \bm{\tau}_{r} \,  \bm{\mathcal{T}}^{-1}(b)  \right] = 
  \tr \left[  \bm{\tau}_{r} \right] \, \tr \left[  \bm{\mathcal{T}}(b)  \right]  - \tr \left[ \bm{\tau}_{r-1}\right] \ .
\end{equation}
Denoting by $z_r = \tr \left[ \bm{\tau}_r(\omega) \right] /2$, then 
\begin{equation}
z_{r+1} = 2 \, z_\infty \, z_r  - z_{r-1}  \ , 
\label{eq045}
\end{equation}
where $z_\infty = \tr \left[\bm{\mathcal{T}}(b) \right] /2 = \lim_{r \to \infty} \tr \left[ \bm{\tau}_r \right]/2$  stands for the spectrum of the system associated to $b$. The two eigenvalues of the recursive scheme \eqref{eq045} are
\begin{equation}
z_\infty \pm  \sqrt{z_\infty^2 - 1}  \ .
\label{eq046}
\end{equation}
 Since the system with transfer matrix $\bm{\mathcal{T}}(b)$ is known a priori, so are its frequency passbands, where $\left|  z_\infty \right|   \leq 1$ holds. In turn, the frequency bands where the latter inequality is verified define the only region where the passbands of the sequence $\{z_r\}$ may converge, according to Eq.~\eqref{eq046}. \\
 
 A closed form expression for $z_{r+1}$ can be determined in this case invoking the property (ii) presented above. Since $\bm{\tau}_{r+1} =\bm{\mathcal{T}}(a) \, \bm{\mathcal{T}}^{r+1}(b)  $, then
 \begin{equation}
 \bm{\tau}_{r+1} =\mathcal{U}_{r}(z_\infty)   \bm{\mathcal{T}}(a) \, \bm{\mathcal{T}}(b) 
 						-\mathcal{U}_{r-1}(z_\infty)   \bm{\mathcal{T}}(a) 
						=\mathcal{U}_{r}(z_\infty)    \bm{\tau}_{1}
						-\mathcal{U}_{r-1}(z_\infty)    \bm{\tau}_{0}		
															\label{eq049}			
 \end{equation}
and taking traces
\begin{equation}
z_{r+1} 						=\mathcal{U}_{r}(z_\infty)    z_1
									-\mathcal{U}_{r-1}(z_\infty)    z_0  \ .
									\label{eq048}
\end{equation}
This expression shows the algebraic pattern of the bands in the $\alpha_r$ sequence, depending on the family of Chevyshev polynomials. The interesting feature is that this structure is repeated independently of the number $a = [0;a_1,\ldots,a_n]$ considered, so that a selfsimilar geometrical shape is to be expected at different scales of the bulk spectrum, something that will be revealed in the numerical experiments in the next section. We shall illustrate the dynamical properties of systems generated by Sturmian sequences using for that both discrete and continuous structures and considering waves of different nature.

\section{Numerical Examples}
\subsection{Example 1. Compressional waves in discrete systems}

\begin{figure}[ht]%
	\begin{center}
		\includegraphics[width=15cm]{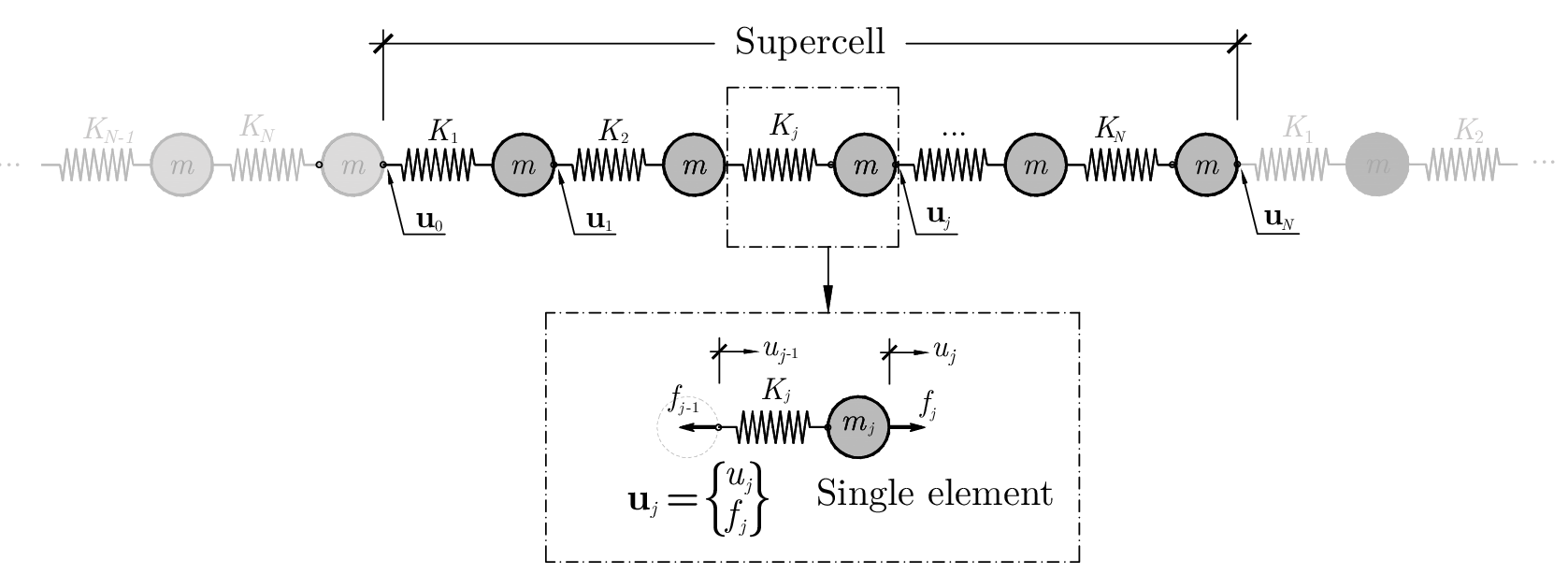} \\	
		\caption{(Example 1) Discrete spring-mass system with Sturmian quasiperiodic distribution of rigidities $K_j$.}%
		\label{fig05}%
	\end{center}
\end{figure}
\begin{figure}[ht]%
	\begin{center}
		\includegraphics[width=17cm]{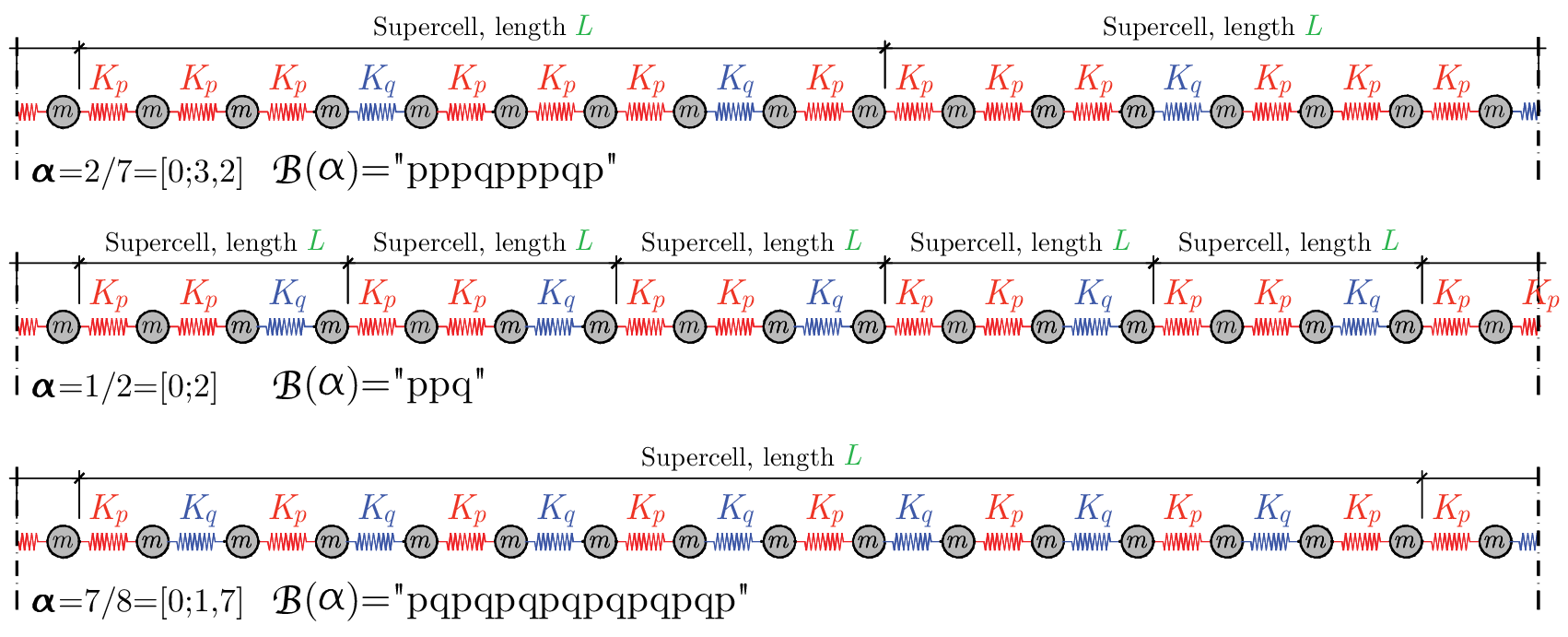} \\	
		\caption{(Example 1) Three spring-mass systems associated to three numbers $\alpha = \{2/7, \ 1/2, \ 7/8\}$. The dispersion relation is determined assuming that the supercell of length $L$ is distributed periodically. For each system $\alpha = \nu_n/\delta_n$ depicts the ratio between the number of rigidities $K_q$ and $K_p$ in each supercell. The size of the system is equal to $\mathcal{N}(\alpha) = \nu_n + \delta_n$.}%
		\label{fig06}%
	\end{center}
\end{figure}

In this first example a discrete spring-mass lattice is considered (see Fig.~\ref{fig05}). Following the methodology described above, the system consists of the periodic concatenation of $N$ single elements formed by a mass and a linear spring, with parameters $m_j$ and $K_j$, respectively. In this example, the rigidities $K_j$ are arranged along the chain following the Sturmian sequence associated to $\alpha = [0;a_1,\ldots,a_n] \in [0,1]$. If $\alpha = \nu_n / \delta_n$, then the amount of entries in the sequence, i.e. the size of the chain, is equal to $\mathcal{N}(\alpha) = \nu_n + \delta_n$. 
The variables $\mathfrak{u}_j(t)$ stand for the horizontal displacements in time domain. Under harmonic motion with circular frequency $\omega$, each degree of freedom can be expressed as $\mathfrak{u}_j(t) = u_j e^{\text{i}\omega t}$ and the harmonic force acting on the link elements is denoted by $\mathfrak{f}_j(t) = f_j e^{\text{i}\omega t}$. Thus, the state vector in the frequency domain can be defined as $\mathbf{u}_{j} = \{u_{j}, f_{j}\}^T$. As shown in Fig.~\ref{fig05}, state vectors can be located at the both ends of each element. The relationship between each state vector and the preceding one is given by the product of the respective transfer matrices associated to the mass and to the spring~\cite{Rui-2019}, i.e.

\begin{multline}
\mathbf{u}_{j} =
\left\{ 
\begin{array}{c}
u_{j} \\
f_{j}
\end{array}
\right\} 
=
\left[ 
\begin{array}{cc}
1 		&		0 \\
-m_j \omega^2 & 1
\end{array} \right] 
\left[ 
\begin{array}{cc}
1 & - \frac{1}{K_{j}} \\
0 & 1
\end{array} \right] 
\left\{ 
\begin{array}{c}
u_{j-1} \\
f_{j-1}
\end{array}
\right\} 
\\
= 
\left[ 
\begin{array}{cc}
1 &  \frac{1}{K_{j}} \\
-m_j \omega^2 & 1 - \frac{ m_j\omega^2}{K_{j}}
\end{array} \right] 
\left\{ 
\begin{array}{c}
u_{j-1} \\
f_{j-1}
\end{array}
\right\} 
\, \equiv \, \mathbf{T}(m_j,K_j) \, \mathbf{u}_{j-1}  \ , 
\label{eq030}
\end{multline}
where $\mathbf{T}(m_j,K_j)$ denotes the transfer matrix of the $j$th element. In this notation is highlighted the fact that $m_j$ and $K_j$ are the dynamical parameters. The generation method considers a certain dynamic property of the system,  $\Theta(j)$, which in this example can be assumed equal to the masses or to the rigidities. In our particular case, we will consider it equal to the stiffnesses, $\Theta(j) = K_j$, remaining constant the masses, i.e. $m_j = m$ for all $j$. The parameter $K_j$ takes values from the binary set $\{K_p,K_q\}$ according to what is specified in the Sturmian block $\mathcal{B}(\alpha)$, which in turn results a binary word from the alphabet $\{p,q\}$. As an example, we shall generate three systems associated with the numbers $\alpha = \{2/7, 1/2, 7/8 \}$ and determine the dispersion relations as well as the representation of the wave frequency bands (spectrum bands).  In the Fig.~\eqref{fig06} the Sturmian blocks and the respective supercells have been sketched. 

\begin{figure}[ht]%
	\begin{flushleft}
		\includegraphics[width=18cm]{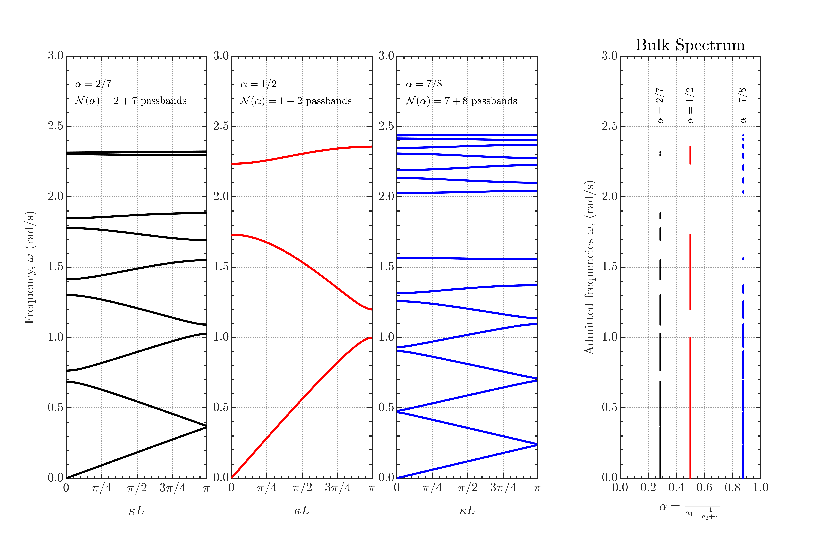} 
		\caption{Example 1. Dispersion relation for three Sturmian spring-mass chains for $\alpha = \{2/7, \ 1/2, \ 7/8\}$. The last figure on the right represents the associated spectrum only for the three numbers considered. }%
		\label{fig07}%
	\end{flushleft}
\end{figure}

The dispersion relation establishes the frequency $\omega$ of the waves admitted in the system associated to the wavenumber $\kappa$, with $\omega$ and $\kappa$ real numbers. The transfer matrix is obtained by the procedure described above in Eqs.~\eqref{eq019} and \eqref{eq020} with
\begin{equation}
\mathbf{T}_q = \mathbf{T}(m,K_q) \quad , \quad 
\mathbf{T}_p = \mathbf{T}(m,K_p)  \ .
\end{equation}
Although omitted, both matrices depend on the frequency. Since spring-mass lattices can be modeled with $2 \times 2$ transfer matrices, the wavenumber $\kappa$ can be determined straightforward from the relationship 
\begin{equation}
\cos (\kappa L) = \frac{1}{2} \, \tr \left[ \bm{\mathcal{T}}(\alpha)  \right]
\label{eq050}
\end{equation}
Using  the particular values $K_p = 1$ N/m, $K_q = 2 K_p = 2 $ N/m, and $m = 1$ kg, the dispersion curves can be constructed sweeping out the range of frequencies $0 \leq \omega \leq 3$ rad/s and solving the above equation for the dimensionless wavenumber $\kappa L$, where $L$ stands for the supercell length. Results are shown in Fig.~\ref{fig07} for the three values $\alpha = \nu_n / \delta_n =  \{2/7, 1/2, 7/8 \}$. The fact that the supercell is made up of single elements with different parameters leads to heterogeneity and therefore to the appearance of passbands and stopbands. It turns out \cite{Hussein-2017,Macia-2009}
%
%
 that the amount of passbands, coincides with the size of the Sturmian sequence which in turn is closely related to the associated number $\alpha$ by $N = \mathcal{N}(\alpha) = \nu_n + \delta_n$.  
Projection on a vertical line of the whole set of admitted frequencies leads to a simplified representation of passbands and stopbands, resulting a dashed line that can be associated to the number $\alpha$, generator of the chain. In Fig.~\ref{fig07} (right) the three numbers are represented in three different colors and their respective spectra have been located on the corresponding abscissas. The graphical representation that arises from the repetition of the process over the entire interval $0 \leq \alpha \leq 1$ gives rise to a figure like that of fig.~\ref{fig09}. We call this representation the {\em Sturmian bulk spectrum}. \\

One of the main objectives of this paper is to determine the Sturmian spectrum for elastic structures and to justify its fractal nature. We shall validate the selfsimilar  properties discussed in the previous section. For this purpose we will consider the quasiperiodic system associated to the number $1/\sqrt{2}$, written as the continuous fraction
\begin{equation}
\frac{1}{\sqrt{2}} = \left[0;1,2,2,2,\ldots\right] = 0.70710678\ldots
\label{eq051}
\end{equation} 
Consider for example its approximation by convergents up to the fourth term, i.e. $a = [0;1,2,2,2] = 12/17 \approx  0.70588$. This number allows to build a sequence given by $\{\alpha_r = [0;1,2,2,2+r]\}_{r=0}^{\infty}$  so that, according to Eqs.~\eqref{eq052} and~\eqref{eq053}, the numbers  $a = \alpha_0 = 12/17$  and $b = \alpha_{\infty} = 5/7 \approx 0.71428$ are the two endpoints of the sequence. The theoretical developments have shown two interesting properties in reference to the spectrum passbands of the systems associated to each $\alpha_r$: (i) the higher the index $r$, the closer the $\alpha_r$ passbands are to those of $b = \alpha_\infty$ and (ii) According to Eq.~\eqref{eq048}, the spectrum of $\alpha_r$ is explicitly defined in terms of the $r$th order Chebyshev polinomial. Therefore,  as $r$ increases, so it does the number of passbands associated to $\alpha_r$. 
\begin{figure}[ht]%
	\begin{center}
			\begin{tabular}{c}
	   \begin{tabular}{rrrc}
	$\alpha_r$				& 		 	$a = \alpha_0$			   & 		$b = \alpha_{\infty}$	&  	Plot  \\
	\hline
	$[0;1,2+r]$				&		2/3 = 0.666667	 &			1			&			\textbf{(a)}   \\
	$[0;1,2,2+r]$			&		5/7 = 0.714285    &     2/3 = 0.666667  &			\textbf{(b)}   \\
	$[0;1,2,2,2+r]$			&      12/17 = 0.705888 &   5/7 = 0.714285    &			\textbf{(c)}   \\
	$[0;1,2,2,2,2+r]$			&      29/41 = 0.707317 &   12/17 = 0.705888  &		\textbf{(d)}   \\
	\hline
\end{tabular} \\				
		\includegraphics[width=17cm]{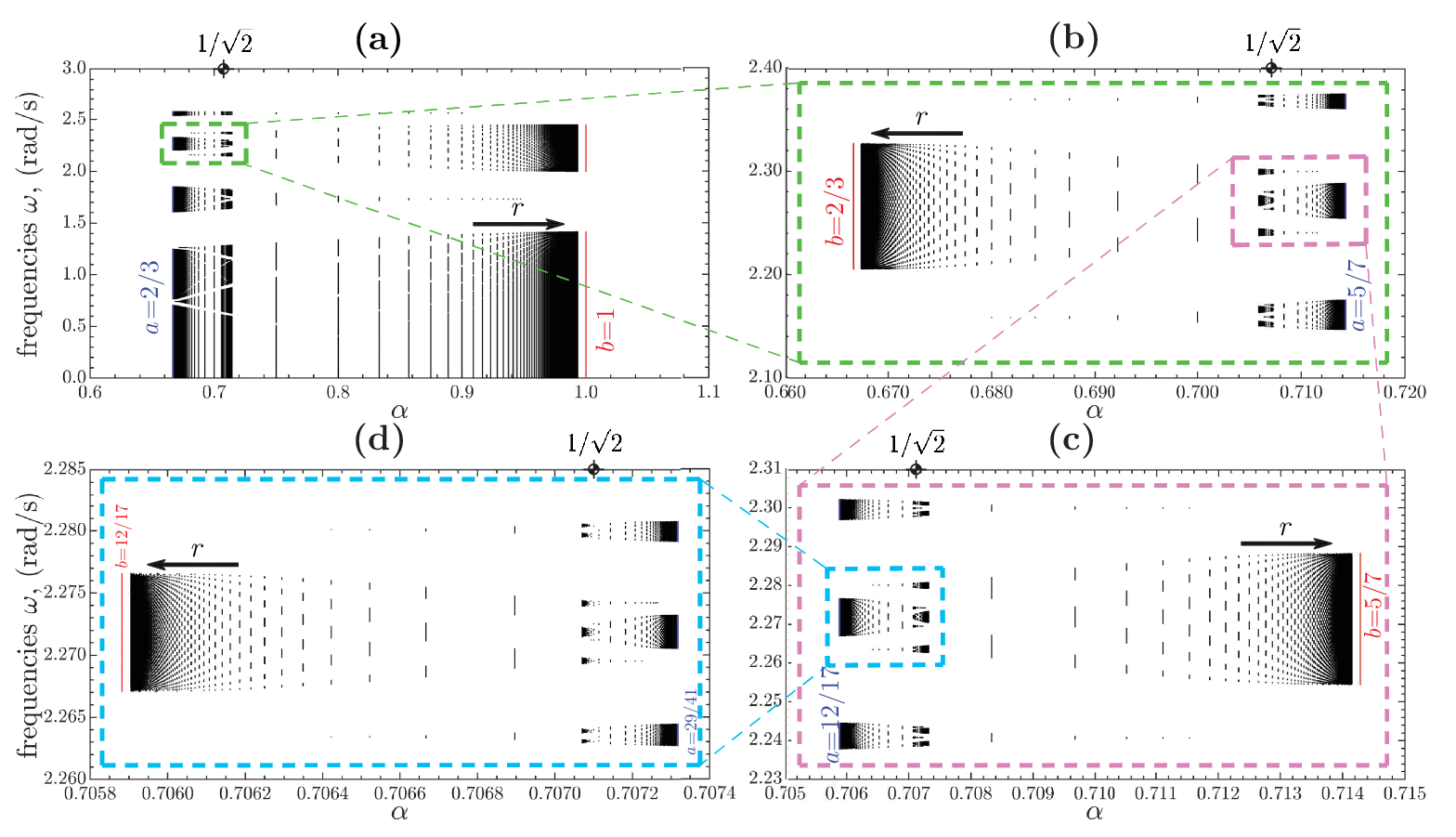} 
			\end{tabular}
		\caption{Sturmian bulk spectrum of  spring-mass systems associated to different sequences $\alpha_r$. The numbers which define the endpoints of the sequence $a=\alpha_0$ and $b =\alpha_\infty$ are listed in the table (above). In this figure, in the upper part, the location of the number $1/\sqrt{2}$ is highlighted. The black arrow indicates the increasing values of $r$. }%
		\label{fig08}%
	\end{center}
\end{figure}
The process described above can be repeated for each of the convergents of the irrational number $1/\sqrt{2}$. We can then distinguish a set of sequences of passbands increasingly closer to the quasiperiodic system defined by the number $1/\sqrt{2}$. Fig.~\ref{fig08} depicts the Sturmian bulk spectrum of the four sequences listed in the table. Self-similar structures formed by the frequency bands arise, highlighting the geometrical meaning of Eq.~\eqref{eq048}, valid regardless of how close the $a=\alpha_0$ and $b=\alpha_\infty$ values are. In the limit as we increase the number of convergents, the same fractal structure appears when zooming around the irrational number $1 / \sqrt{2}$. 
\\

\begin{figure}[h!]%
	\begin{center}
		\begin{tabular}{cc}
					\includegraphics[width=8.5cm]{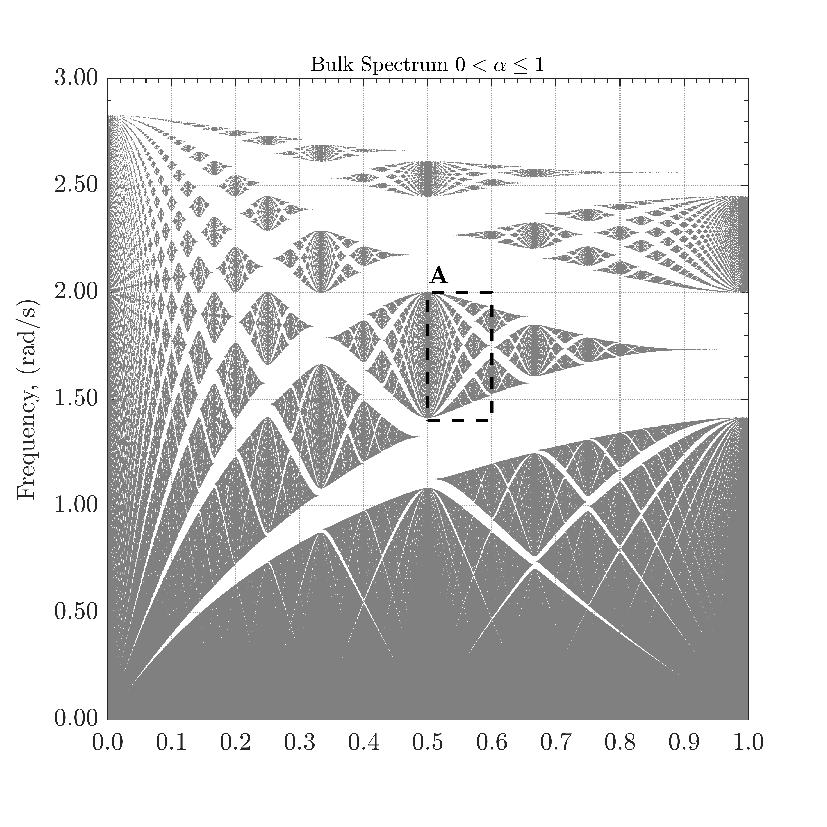}  & 
					\includegraphics[width=8.5cm]{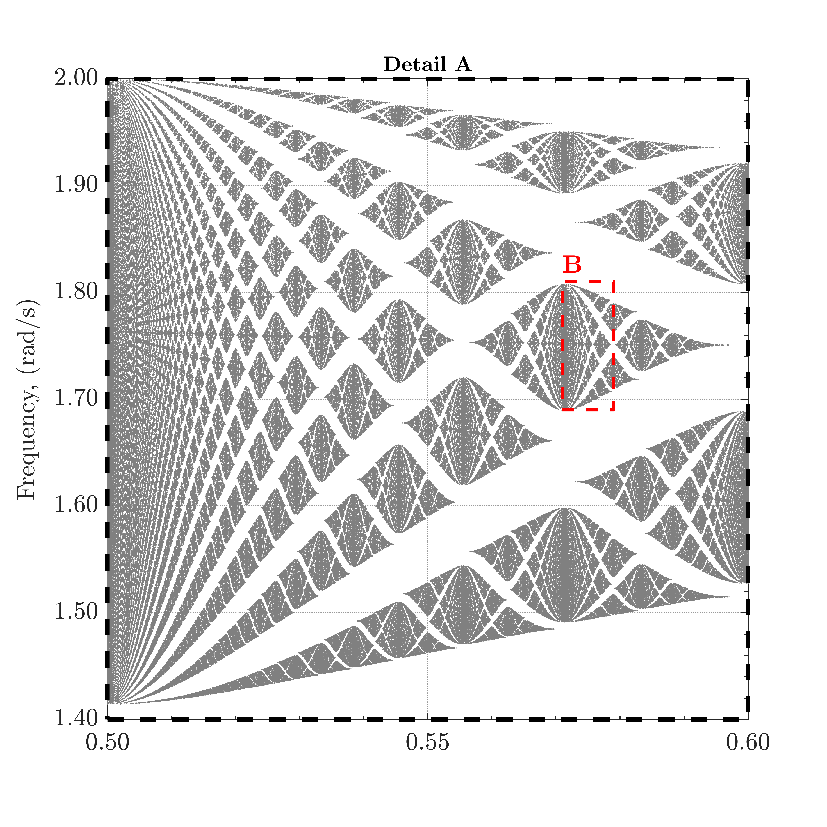}  \\
					\includegraphics[width=8.5cm]{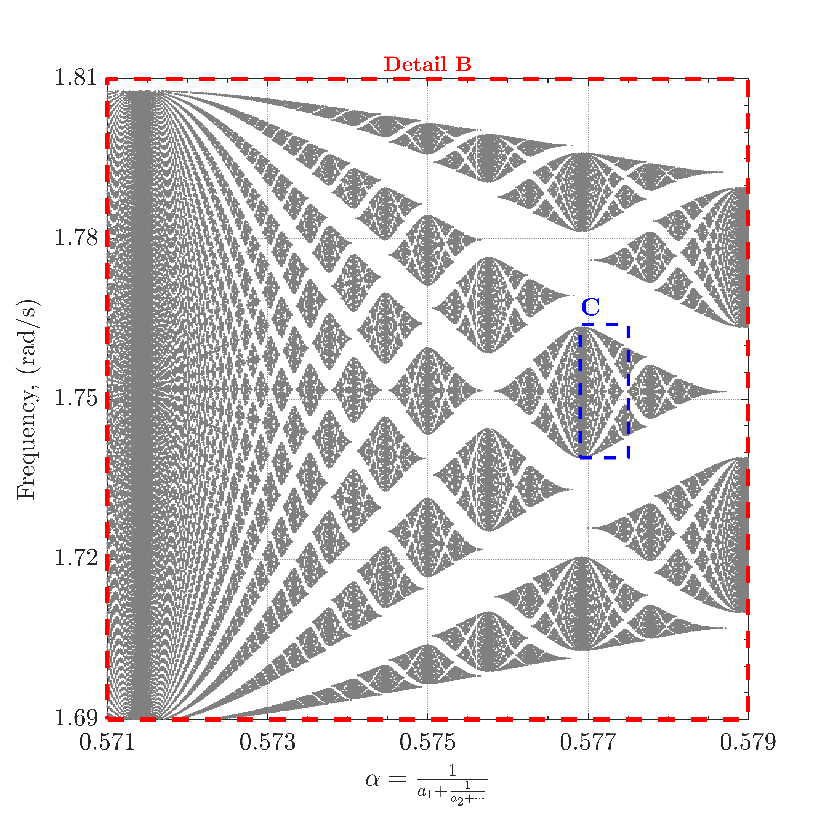}  & 
					\includegraphics[width=8.5cm]{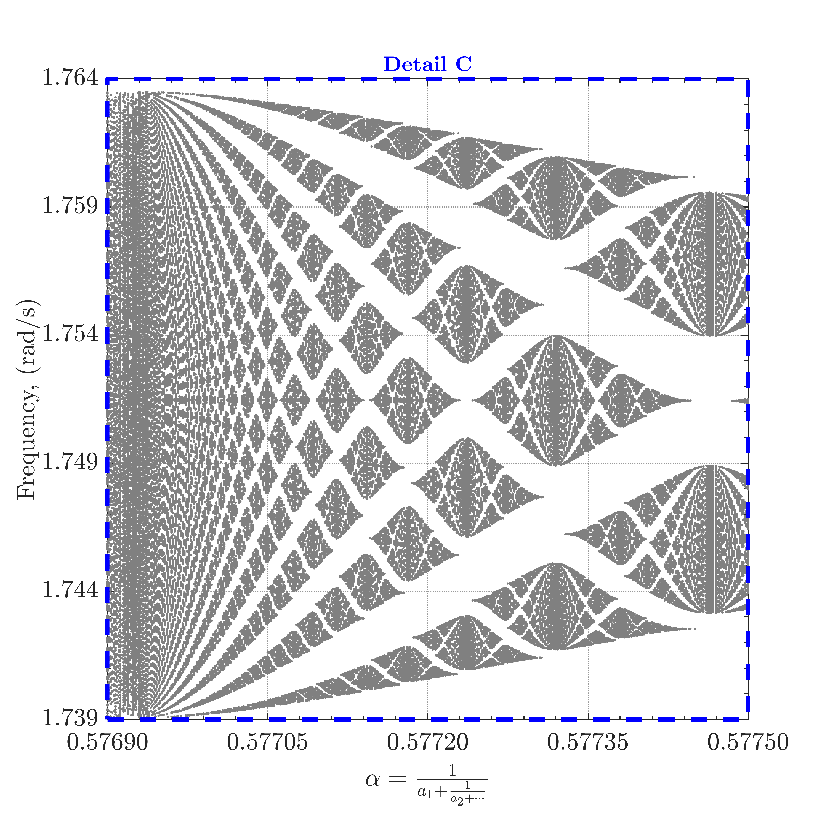}  \\										
		\end{tabular}
		\caption{Sturmian bulk spectrum of a spring-mass system with quasiperiodic distribution of rigidities $K_j$. Top-left: bulk spectrum for the whole range of generator parameter $0\leq \alpha \leq 1$. Top-right, bottom-right and bottom-left: details A, B and C to visualize the selfsimilar structure of the bulk spectrum. }%
		\label{fig09}%
	\end{center}
\end{figure}

In fig.~\ref{fig09} the Sturmian bulk spectrum in the interval $0\leq \alpha \leq 1$ for the parameters $K_p = 1$ N/m and $K_q = 2K_p = 2 $ N/m and $m = 1$ kg has been plotted. The fractal pattern observed above in fig.~\ref{fig08} is now reproduced revealing more complex shapes. One can easily distinguish the simplest systems, those generated by simple fractions of the form $\alpha = \{1, 1/2, 1/3, 1/4, 1/5,\ldots\}$ and $\alpha = \{2/3,  3/4,   4/5,\ldots\}$ because they have the widest frequency bands, shown darker in the figure. In particular $\alpha=0$ corresponds to the degenerate system ``$qqqqq\ldots$'', while $\alpha=1$ gives rise to the periodic binary system ``$pqpqpq \ldots$'', with two bands: the low frequency or acoustic band and the high frequency or optical band. Both $\alpha=0,1$ have the well--known $\kappa$--$\omega$ spectrum given by expressions\\
\begin{eqnarray}
\alpha=0 & \to & \cos (\kappa L) = 1 - \frac{\omega^2}{2 K_q}  \ ,  \nonumber \\  
\alpha=1 & \to & \cos (\kappa L) = 1 - \frac{K_p+K_q}{K_p K_q} \omega^2 + \frac{\omega^4}{2 K_p K_q}  \ .\label{eq070} 
\end{eqnarray}
Both 0 and 1 are precisely the limits of two sequences of type $\alpha_r$ that begin at the same point $\alpha_0 = 1/2$. Indeed, $\alpha_r = [0;2+r] = \{1/2,1/3,1/4,\ldots\}$ has as limit $\alpha_\infty = 0$, while $\alpha_r = [0;1,1+r]=\{1/2,2/3,3/4,\ldots\}$ converges to $\alpha_\infty = 1$. The figure~\ref{fig09} clearly reveals the bands associated with these two sequences, which are in turn limits of other sequences of numbers with increasingly longer continued fractions.

\subsection{Example 2. Compressional  waves in rods}

\begin{figure}[ht]%
	\begin{center}
		\begin{tabular}{ccc}
		\multicolumn{3}{c}{\includegraphics[width=14cm]{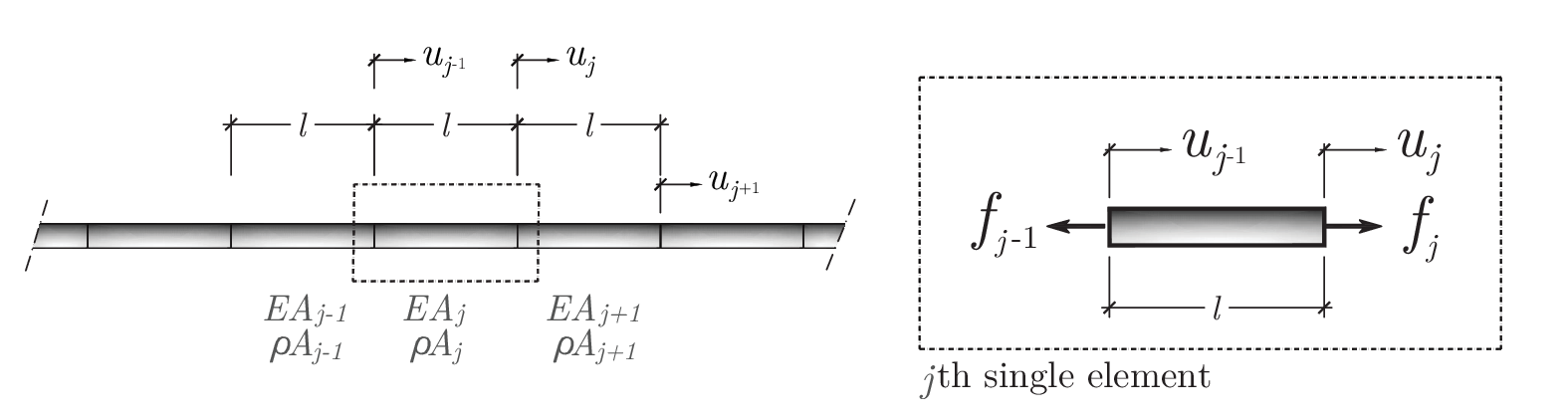}}		  \\
		\includegraphics[width=5.5cm]{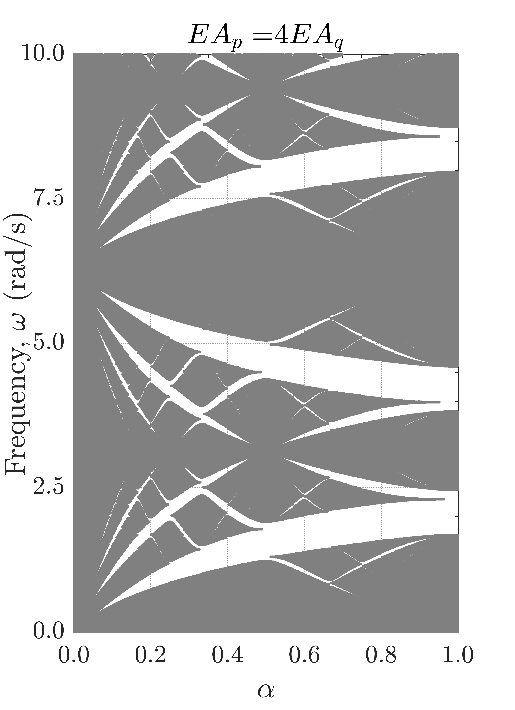}          &
		\includegraphics[width=5.5cm]{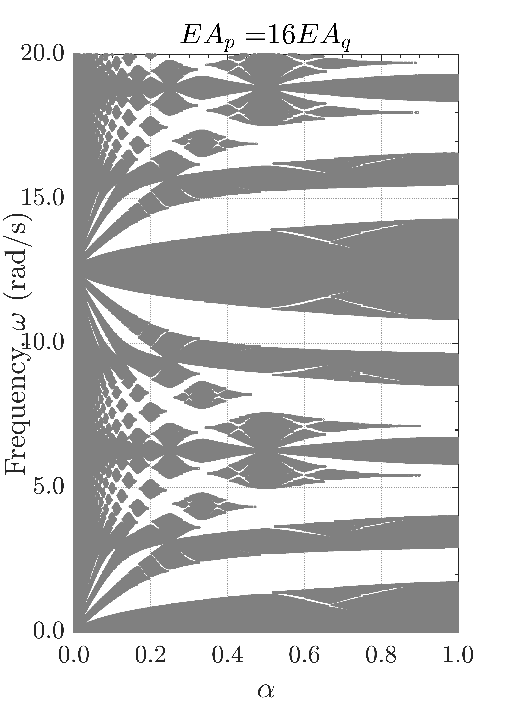}		&
		\includegraphics[width=5.5cm]{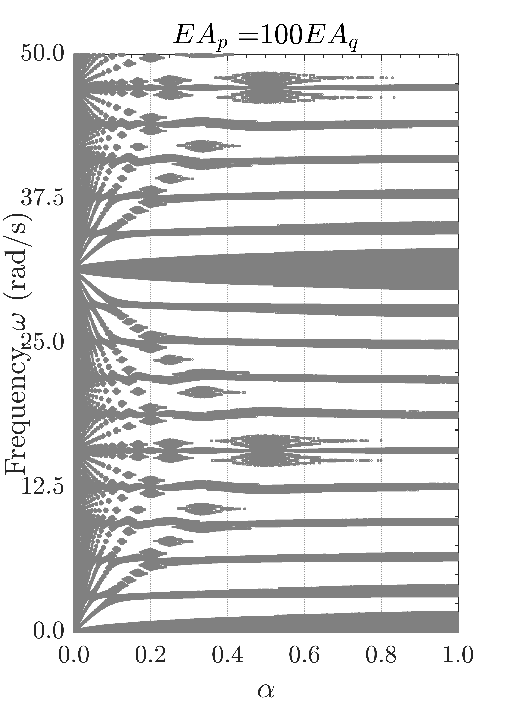}		\\
		\includegraphics[width=5.5cm]{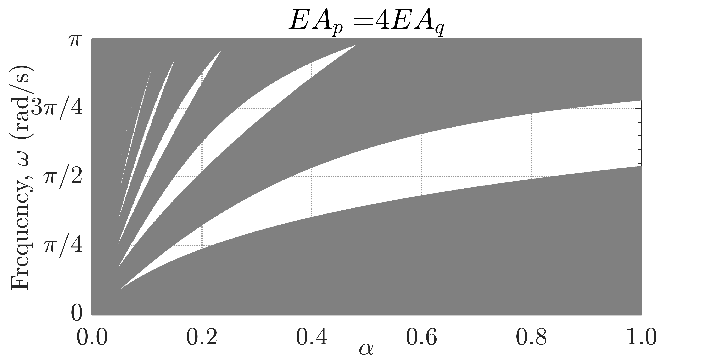}        &
		\includegraphics[width=5.5cm]{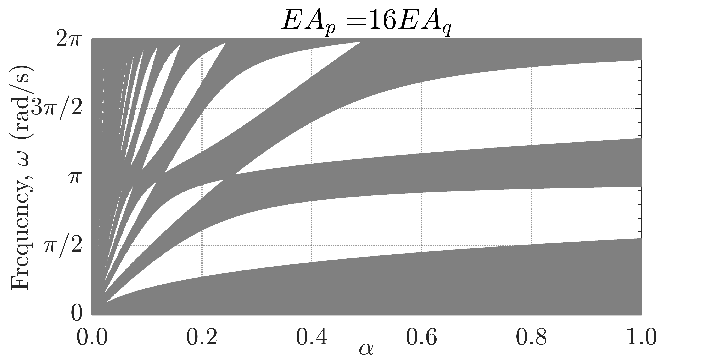}		&
		\includegraphics[width=5.5cm]{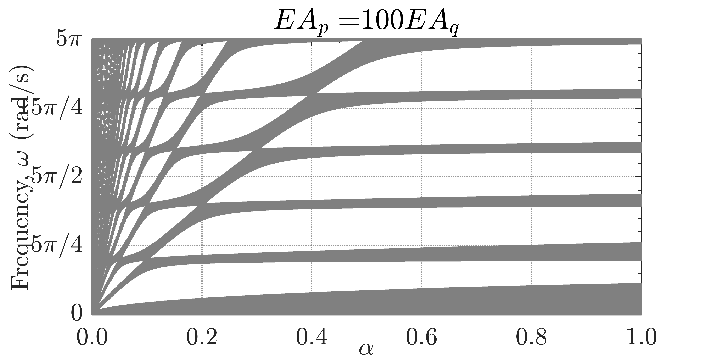}				
		\end{tabular}
		\caption{Sturmian bulk spectra of a rod (compressional waves) with quasiperiodic variation of elastic sectional stiffness between values $\{EA_p,EA_q\}$: Left, middle and right plots show the bulk spectrum for three different values of the ratio $EA_p/EA_q$: $EA_p = 4 EA_q$ (left), $EA_p = 16 EA_q$ (middle) and $EA_p = 100 EA_q$ (right). Darkened regions show frequency passbands. The three plots on bottom represent the frequency passbands obtained from the analytical expression~\eqref{eq058}, for the three values of $EA_p/EA_q$ considered}%
		\label{fig10}%
	\end{center}
\end{figure}

In this example Sturmian distribution of parameters along a straight rod will be considered. Let us assume an infinite medium formed by single elements of length $l$. The $j$th element has stiffness and mass properties given by $EA_j$ and $\rho A_j$, where $EA_j$ and $\rho A_j$ stand for the compressional sectional stiffness and the mass per unit of length, respectively.
In order to simplify the notation, the parameters $EA_j$ and $\rho A_j$ are understood as the products of the Young modulus $E_j$ and the density $\rho_j$ and the area of the cross section $A_j$, associated to the $j$th element.  As known, horizontal displacement $u(x,t)$ and internal force $f(x,t)$ in the $j$th element are related by
\begin{equation}
\frac{\partial u}{\partial x} = \frac{f(x,t)}{EA_j} \quad , \quad 
 \frac{\partial f}{\partial x} = \rho A_j \frac{\partial^2 u}{\partial t^2}  \ .
 \label{eq054}
\end{equation}
Assuming harmonic motion with $u(x,t) = U(x) \, e^{\text{i}\omega t}$ and $f(x,t) = F(x) \, e^{\text{i}\omega t}$, Eqs.~\eqref{eq054} yields
\begin{equation}
\left\lbrace 
\begin{array}{c}
U'(x) \\
F'(x) 
\end{array}
\right\rbrace 
=
\left[ 
\begin{array}{cc}
0   &   1/EA_j   \\
-\omega^2 \, \rho A_j   &  0 
\end{array}\right] 
\left\lbrace 
\begin{array}{c}
U(x) \\
F(x) 
\end{array}
\right\rbrace   \ , 
\end{equation}
where $(\bullet)' = d(\bullet) / dx$ denotes the space--domain derivative. The state vector $\mathbf{u}(x) = \{U(x), F(x)\}^T$ verifies then $\mathbf{u}' = \mathbf{W}_j(\omega) \, \mathbf{u}$, which integrating between $x=0$ and $x=l$ give rise to the transfer matrix of a single element, $\mathbf{u}(l) = e^{\mathbf{W}_j(\omega)l} \, \mathbf{u}(0)$, where
\begin{equation}
\mathbf{T}_j(\omega) = e^{\mathbf{W}_j(\omega)l} = 
\left[ 
\begin{array}{cc}
\cos \mu_j 		&  \frac{l }{EA_j \, \mu_j  	} \sin \mu_j \\
 - \mu_j	 \frac{EA_j}{l} \, \sin \mu_j   &   \cos \mu_j 
\end{array}
\right] \quad , \quad 
\mu_j = \omega \, l \, \sqrt{\frac{\rho A_j}{EA_j }}   \ .
\end{equation} 
Therefore, the relationship between both state vectors $\mathbf{u}_j = \mathbf{u}(l)$ and $\mathbf{u}_{j-1} = \mathbf{u}(0)$  has been established as function of the transfer matrix. The algorithm of the Sturmian sequences can already be applied in order to generate the bulk-spectrum of compressional waves traveling through rod structures. In this case, we consider that the sectional stiffness $EA$ takes the role of the $\Theta$ parameter used in the theoretical developments. Thus, $EA_j \in \{EA_p, EA_q\}$. To carry out the numerical computations we subdivide the interval $[0,1]$ into 1000 points and evaluate in a loop the transfer matrix $\bm{\mathcal{T}}(\alpha)$ for each point. Thus,  we can then aim to determine the dispersion relations associated with numbers of at most 3 decimal places. As can be seen from the numerical results in Fig. \ref{fig10}, it is more than enough to notice the resulting geometrical pattern. We know that $\alpha=0$ corresponds to the continuous homogeneous rod with parameters $EA_q$ and $\rho A_q$ with no stopbands in the whole frequency band. On the other side, $\alpha=1$ gives rise to the periodic binary system ``$pqpqpq \ldots$''.
In fig.~\ref{fig10} the bulk spectrum for different ratios $EA_p/EA_q$ have been represented. Since rods are continuous structures, we will find  passbands in the whole frequency band. However, as observed in the three plots, the general pattern of bands distribution strongly depends on the contrast between both $EA_p$ and $EA_q$. Furthemore, a periodicity is observed in the vertical direction (frequencies) of the bulk spectrum. Both, the bands width and the periodicity can be explained and somehow quantified studying the spectrum of the systems associated to the numbers given by the sequence $\{\alpha_r = 1/r\}_{r=1}^\infty$. The associated Sturmian block  is 
$\mathcal{B}(\alpha_r) = p \stackrel{r}{\ldots}  p  \, q$ and therefore the transfer matrix yields
\begin{equation}
\bm{\mathcal{T}}(\alpha_r) = \mathbf{T}_q(\omega) \,  \mathbf{T}_p^r(\omega)  \ , 
\label{eq055}
\end{equation}
where
\begin{eqnarray}
\mathbf{T}_q(\omega) &=& e^{\mathbf{W}_q(\omega)l} = 
\left[ 
\begin{array}{cc}
\cos \mu_q 		&  \frac{l }{EA_q \, \mu_q  	} \sin \mu_q \\
- \mu_q	 \frac{EA_q}{l} \, \sin \mu_q   &   \cos \mu_q 
\end{array}
\right] \quad , \quad 
\mu_q = \omega \, l \, \sqrt{\frac{\rho A_q}{EA_q }}   \ , \nonumber \\
\mathbf{T}^r_q(\omega) &=& e^{\mathbf{W}_p(\omega) (rl)} = 
\left[ 
\begin{array}{cc}
\cos (r\mu_p) 		&  \frac{l }{EA_p \,  \mu_p  	} \sin (r\mu_p) \\
- \mu_p	 \frac{EA_p}{l} \, \sin (r \mu_p)   &   \cos (r\mu_p) 
\end{array}
\right] \quad , \quad 
\mu_p = \omega \, l \, \sqrt{\frac{\rho A_p}{EA_p }}   \ .\nonumber \\
\end{eqnarray} 
The spectrum of admitted frequencies can be found as the values of $\omega \in \mathbb{R}$ such that $-1 \leq z_r(\omega) \leq 1$, where $z_r(\omega)$ stands for the half trace of the transfer matrix, which after some simplifications can be expressed as
\begin{eqnarray}
z_r(\omega) &=& \frac{1}{2} \tr \left[ \mathbf{T}_q(\omega) \mathbf{T}_p^r(\omega) \right] \nonumber \\
					&=& 
					\frac{(1 + \lambda)^2}{4\lambda} \, \cos \left( \frac{\lambda + r}{\lambda} \, \frac{\omega}{l c_q} \right)  - 
					\frac{(1 - \lambda)^2}{4\lambda} \, \cos \left( \frac{\lambda - r}{\lambda} \, \frac{\omega}{l c_q} \right) 
					\ , \quad \lambda = \sqrt{\frac{EA_p}{EA_q}} \ , \ 
					c_q =  \sqrt{\frac{EA_q}{\rho A_q}}  \ .
					\label{eq056}
\end{eqnarray}

The conditions for the above expression to be periodic in frequency is that $(\lambda + r )/(\lambda - r)$ is rational, something that it holds provided that $\lambda$ is rational. In fig.~\ref{fig10} the bulk spectrum for $\lambda = 2, \ 4, \ 10$ have been plotted. Along the frequency direction, the figures have a periodicity equal to $\Delta \omega = \pi \lambda c_q / l$. The particular values of the parameters are $\rho A_p = \rho A_q = 1$ kg/m, $EA_p = \lambda^2 EA_q$, $EA_q = 1$ N/m, $c_q = 1$ m/s. Therefore, $\Delta \omega = \{2 \pi, 4 \pi, 5 \pi\}$ rad/s. The three plots show clearly the periodicity not only for the those values corresponding to $\alpha_r = 1/r$, but also for the whole range $0\leq \alpha \leq 1$. The higher the ratio $\lambda = \sqrt{EA_p/EA_q}$, the more contrast between both rigidities. It is then expected that the passbands  become narrower, as indeed occurs in the right plot. \\

Finally, we consider of interest to present an analytical outcome which reproduces part of the pattern shown in figs.~\ref{fig10}. A motivation along this paper is to achieve a formula of the spectrum depending analytically on $\alpha$ and $\omega$. The closest we were of such an expression is that one depending on the Chebyshev polynomials derived in Eq.~\eqref{eq048}, valid for $2 \times 2$ transfer matrices. However, this formula does not include the number $\alpha$ explicitly. In the case of rods, the closed form given in Eq.~\eqref{eq056} has been determined. Although it is essentially exact and reproduces passbands and stopbands for all integers  $r = 1 / \alpha_r$. In this paper we wonder how the bands are   distributed if we do $r = 1/\alpha$, allowing $\alpha$ to take any real number in the range $0\leq \alpha \leq 1$, leading to the new formula
\begin{equation}
Z(\alpha,\omega) =
\frac{(1 + \lambda)^2}{4\lambda} \, \cos \left( \frac{\alpha\, \lambda + 1}{\alpha \, \lambda} \, \frac{\omega}{l c_q} \right)  - 
\frac{(1 - \lambda)^2}{4\lambda} \, \cos \left( \frac{\alpha \,\lambda - 1}{\alpha \,\lambda} \, \frac{\omega}{l c_q} \right) 
\label{eq058}
\end{equation}
It is important to note that, although it is a closed form, it is an expression derived after substituting $\alpha_r$ by $\alpha$. Its representation in figs.~\ref{fig10}(bottom) is made in order to experiment numerically what happens and to observe how it reproduces passbands and stopbands. Thus, we see that
\begin{itemize}
	\item The representation of the set $\{(\alpha,\omega): -1 \leq Z(\alpha,\omega) \leq 1, \}$ reproduces the global form of the wider stopbands of the original bulk spectrum, but it does for $0\leq \omega \leq \Delta \omega = \pi \lambda c_q/2l $. Further, the form is completely different. In the fig.~\ref{fig10}, it has only been depicted this range, which covers the half period of the bulk spectrum. 
	\item Eq.~\eqref{eq058} reproduces the width pattern of the passbands: the larger the ratio $\lambda$, the narrower the passbands.
	\item The fractal structure of the spectrum is not replicated. The admitted frequency bands do not show self-similarity.
\end{itemize}
Research on the formula developed and the explanation of the different phenomena observed is left for future work. 

\subsection{Example 3. Flexural waves in beams}

In this last example the  bulk spectrum of Sturmian structured beams will be studied. The Timoshenko beam model, which includes transverse shear deformation and rotational inertia, results of special interest because it can be modeled with $4 \times 4$ transfer matrices. The state variables of any point $x$ along the axis of the beam are one side displacements and rotations $w(x,t)$ and $\varphi(x,t)$ and on the other side, shear force and bending moment (internal forces), i.e. $\mathcal{V}(x,t)$ and $\mathcal{M}(x,t)$.  Constitutive relationships and dynamic equilibrium lead to the four partial differential equations relating these variables~\cite{Doyle-1997}
\begin{eqnarray}
\frac{\partial w}{\partial x} &=& \frac{\mathcal{V}}{GA} + \varphi   \ , \label{eq060c}  \\
\frac{\partial \varphi}{\partial x} &=& \frac{\mathcal{M}}{EI}  \ ,  \label{eq060d}   \\
\frac{\partial \mathcal{V}}{\partial x}   &=& \rho A \, \frac{\partial^2 w}{\partial t^2}  \ , \label{eq060a} \\
\frac{\partial \mathcal{M}}{\partial x} &=& \rho I \, \frac{\partial^2 \varphi}{\partial t^2}  \ , \label{eq060b} 
\end{eqnarray}
where $EI, GA$ stand for the bending and shear stiffnesses, $\rho A, \ \rho I$ are the mass and rotational inertia per unit of length, respectively. Assuming harmonic motion, 
\begin{equation}
w(x,t) = W(x) \, e^{\text{i} \omega t} \ , \quad 
\varphi(x,t) = \phi(x) \, e^{\text{i} \omega t} \ , \quad 
\mathcal{V}(x,t) = V(x) \, e^{\text{i} \omega t} \ , \quad 
\mathcal{M}(x,t) = M(x) \, e^{\text{i} \omega t}  \ , 
\label{eq063}
\end{equation}
the above equations of motion can be reduced to a system of 4 ordinary differential equations in the space domain, yielding in matrix form
\begin{equation}
\frac{\mathrm{d}  \mathbf{u}}{ \mathrm{d} x} = \mathbf{W} \, \mathbf{u}  \ , 
\label{eq062}
\end{equation}
where $\mathbf{u}(x) = \{W(x), \phi(x), V(x), M(x) \}^T$ denotes the state vector (in frequency domain) and 
\begin{equation}
\mathbf{W} = 
\left[ 
\begin{array}{cccc}
0							&		1		&		1/GA		&		0		\\
0							&		0		&				0		&	1/EI		\\
-\rho A \omega^2 & 	    0 	    & 				0 		& 		0  \\ 
0							& -\rho I \omega^2 & 	    0 	    & 				0 	 \\ 
\end{array}
\right]   \ .
\label{eq064}
\end{equation}
The integration of Eq.~\eqref{eq062} along a single element allows us to determine the relationship between the state vectors $\mathbf{u}_j = \mathbf{u}(x_j)$ and $\mathbf{u}_{j-1} = \mathbf{u}(x_{j-1})$, with $l_j = x_j - x_{j-1}$, yielding 
\begin{equation}
 \mathbf{u}_j = \mathbf{T}_j(\omega) \mathbf{u}_{j-1}  \ , 
 \label{eq68}
\end{equation} 
where the transfer matrix of a single element of length $l_j$ is
\begin{equation}
\mathbf{T}_j(\omega) = e^{\mathbf{W}l_j} = 
\left[ 
\begin{array}{cccc}
\cos(\kappa_s l_j)   &
\frac{\kappa_b \sin(\kappa_b l_j) - \kappa_s \sin(\kappa_s l_j) }{\kappa_b^2 - \kappa_s^2}		&
\frac{\kappa_s \sin (\kappa_s l_j) }{\rho A_j \omega^2}	&
\frac{\kappa_b^2 (\cos (\kappa_sl_j) - \cos (\kappa_bl_j)  ) }{\rho I_j \omega^2 (\kappa_b^2 - \kappa_s^2)}			\\
0							&		\cos(\kappa_b l_j)		&				0		&	\frac{\kappa_b \sin (\kappa_b l_j) }{\rho I_j \omega^2}			\\
-\frac{\rho A_j \omega^2 \sin (\kappa_s l_j) }{\kappa_s }	 &
-\frac{\rho I_j \omega^2  (\cos (\kappa_sl_j) - \cos (\kappa_bl_j)  ) }{\kappa_b^2 (\kappa_b^2 - \kappa_s^2)}		 	    &
\cos(\kappa_s l_j)  			&
 \frac{\kappa_b \, \rho A_j}{\kappa_s \, \rho I_j} \frac{\kappa_b \sin(\kappa_b l_j) - \kappa_s \sin(\kappa_s l_j) }{\kappa_b^2 - \kappa_s^2}	  \\ 
0							& - \frac{\rho I_j \omega^2  \sin (\kappa_b l_j) }{\kappa_b}		& 	    0 	    & 				\cos (\kappa_b l_j) 	 \\ 
\end{array}
\right]    , 
\label{eq065}
\end{equation}
where $\kappa_b$ and $\kappa_s$ represent respectively two parameters with wavenumber dimensions given by
%
%
\begin{equation}
\kappa_b = \omega \, \sqrt{\frac{\rho I_j}{EI_j}} \quad , \quad 
\kappa_s = \omega \, \sqrt{\frac{\rho A_j}{GA_j}}  \ .
\label{eq066}
\end{equation}
%
and $\rho A_j, \ \rho I_j, \ EI_j, GA_j$ denote respectively the mass per unit of length, the rotational inertia, the bending stiffness and the shear stiffness of the $j$th single element. In case of Sturmian structured media, one of the above properties is assumed to take two possible values as ruled by the Sturmian sequence. 
\begin{figure}[ht]%
	\begin{center}
		\begin{tabular}{c}
			\includegraphics[width=14cm]{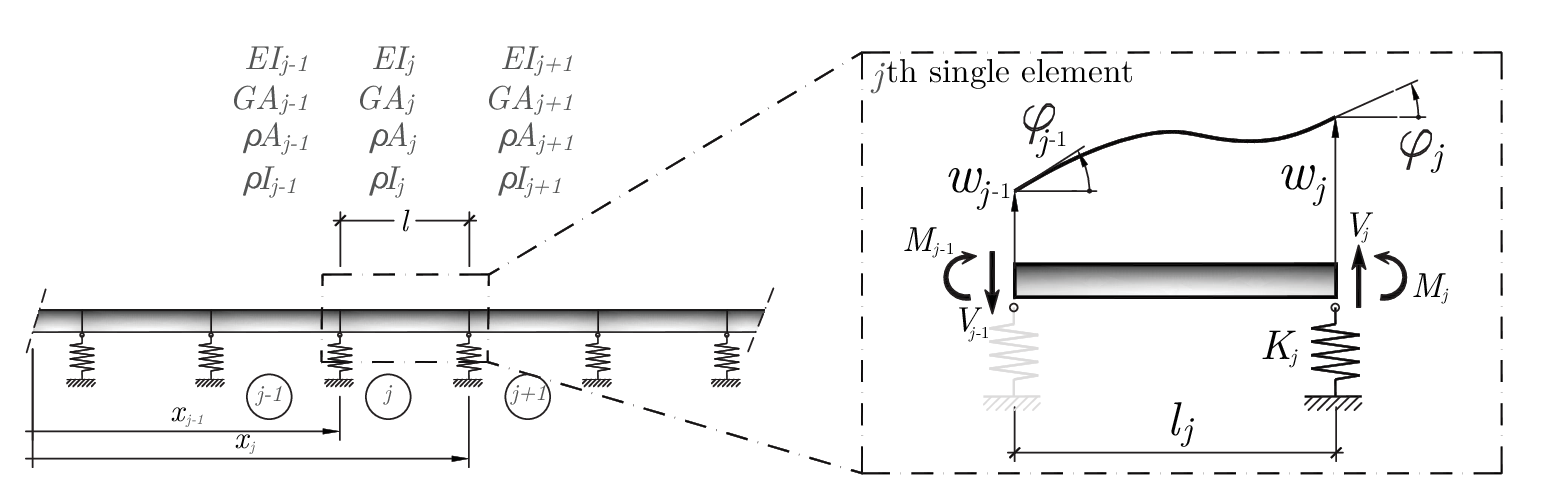} 
		\end{tabular}
		\caption{Quasiperiodic structured beam based on the Timoshenko model. Geometrical and material properties ($EI_j, GA_j, \ldots$) are associated to each element. }%
		\label{fig11}%
	\end{center}
\end{figure}
Furthermore, even more complex systems can be constructed just multiplying by the corresponding transfer matrix. Thus, for instance in fig.~\ref{fig11} a single spring of rigidity $K_j$ has been added. The transfer matrix of the single element is then the product $\mathbf{T}(K_j) \, \mathbf{T}_j(\omega)$ where
\begin{equation}
\mathbf{T}(K) =
\left[ 
\begin{array}{cccc}
1		&		0		&		0		&		0		\\
0		&		1		&		0		&		0		\\
-K   	&		0		&		1		&		0		\\
0		&		0		&		0		&		1		\\
\end{array}
\right]   \ .
\label{eq067}
\end{equation}

\begin{figure}[ht]%
	\begin{center}
		\begin{tabular}{cc}
			\textbf{(a)} \includegraphics[width=4cm]{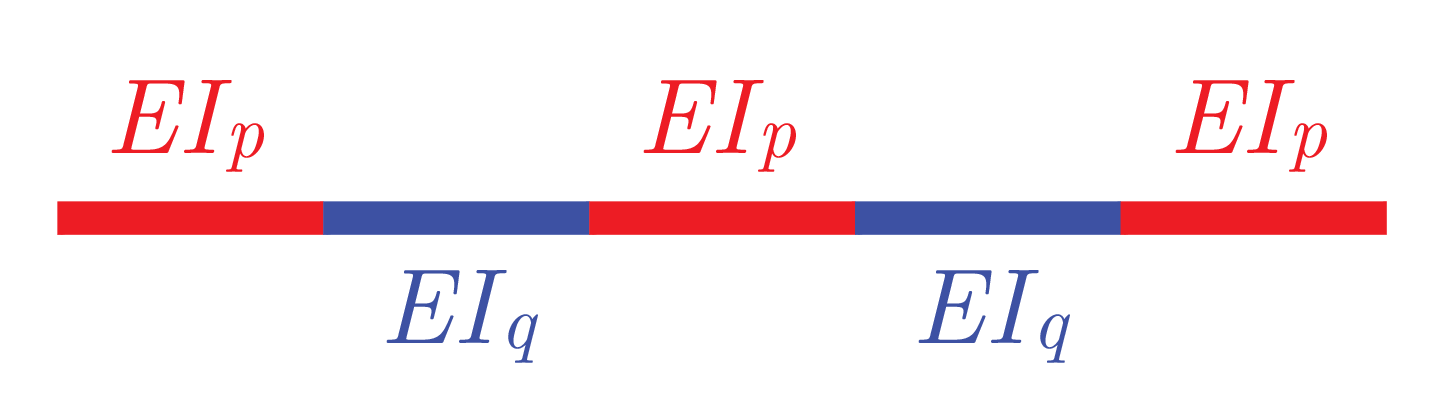}          &
			\textbf{(b)}  \includegraphics[width=4cm]{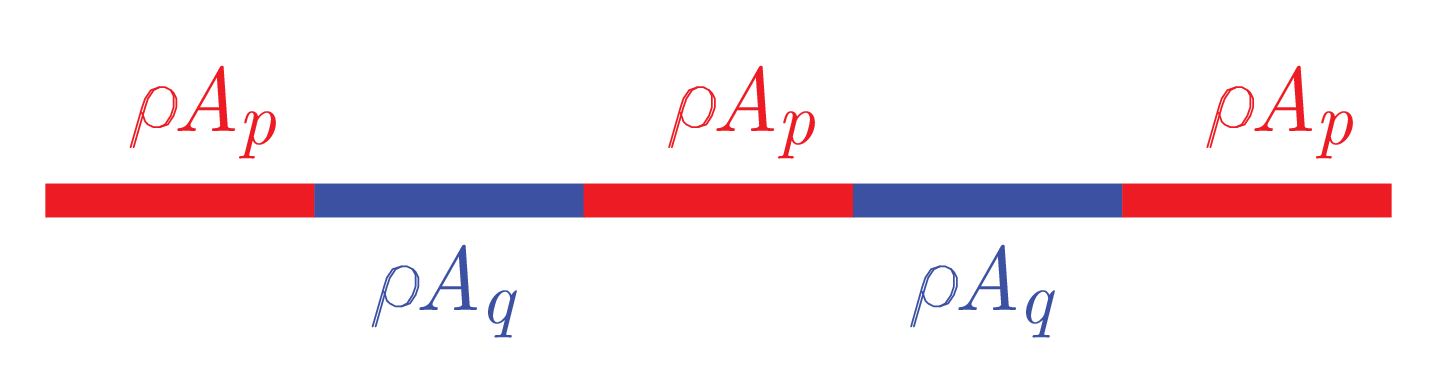}		\\
			\includegraphics[width=8.5cm]{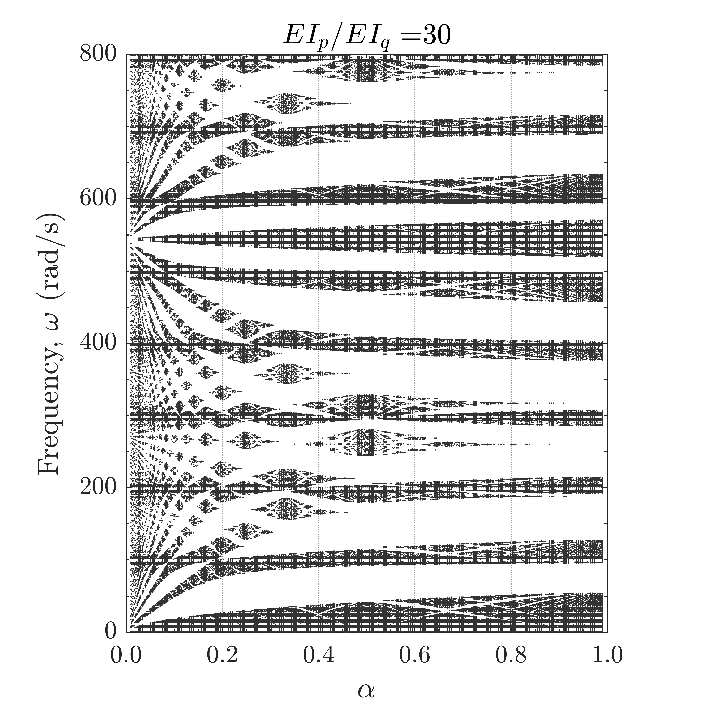}    			&			
			\includegraphics[width=8.5cm]{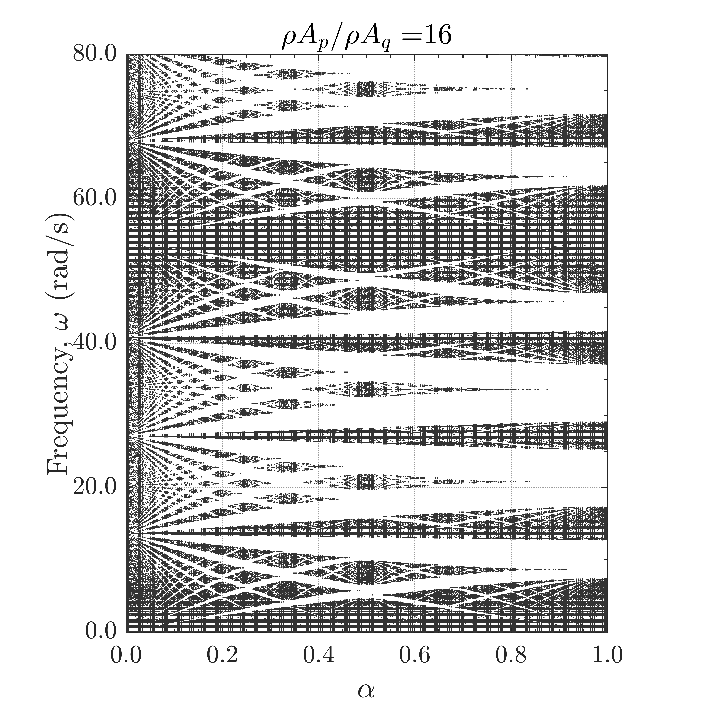}    			\\
			\textbf{(c)} \includegraphics[width=4cm]{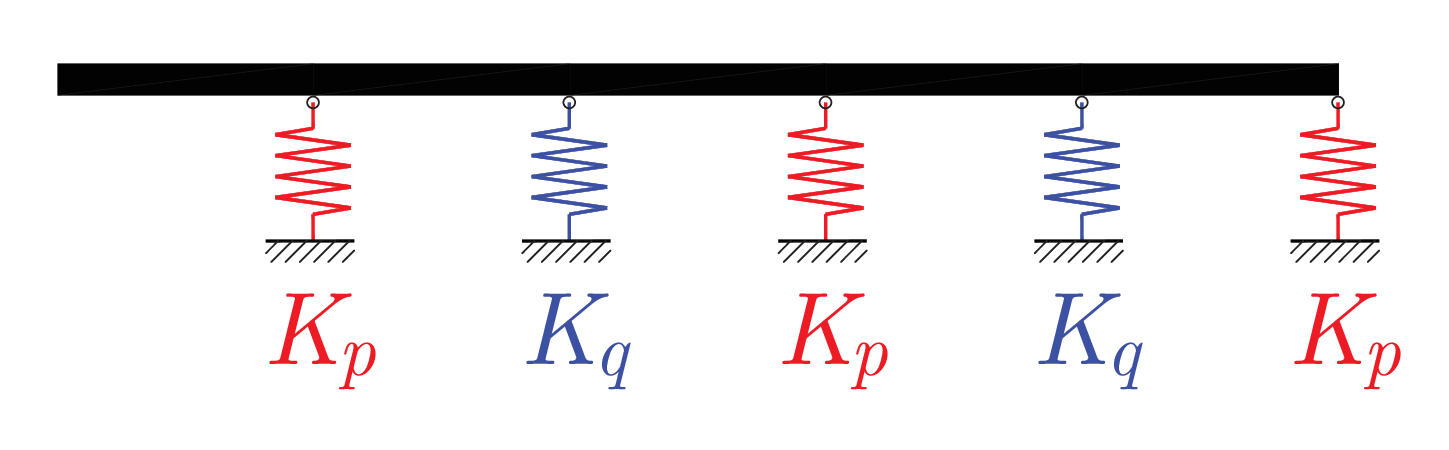}		&
			\textbf{(d)} \includegraphics[width=4cm]{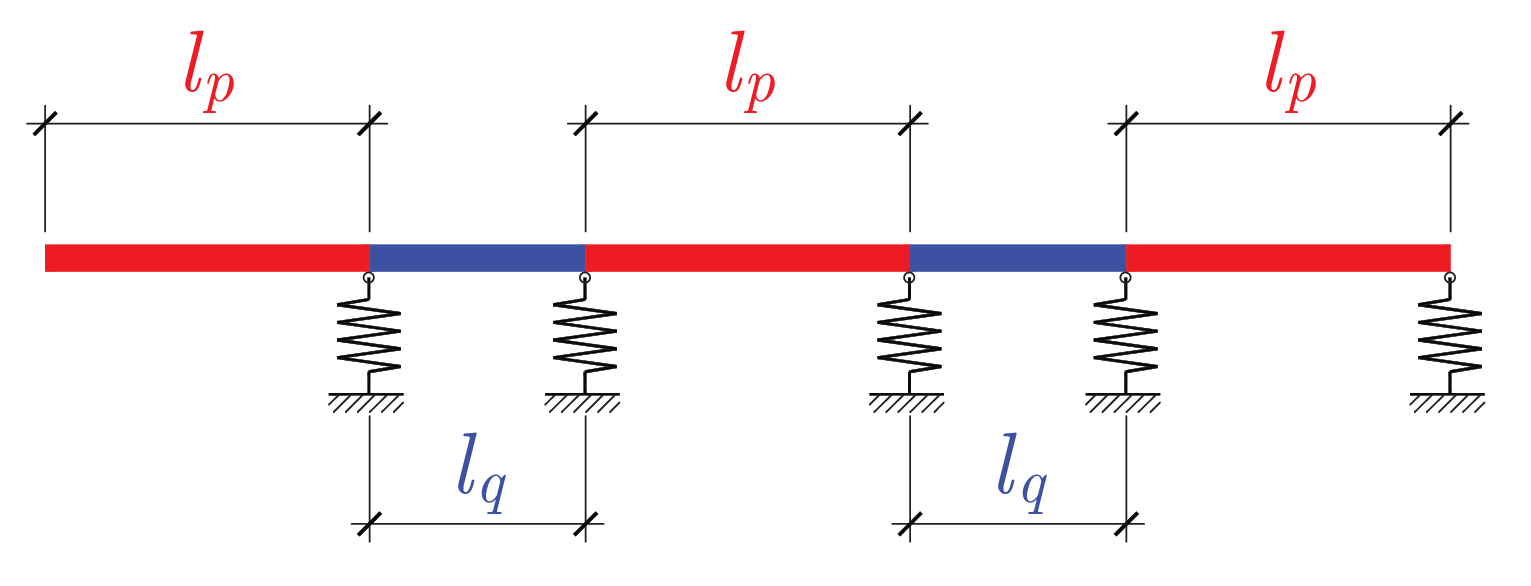}        \\
			\includegraphics[width=8.5cm]{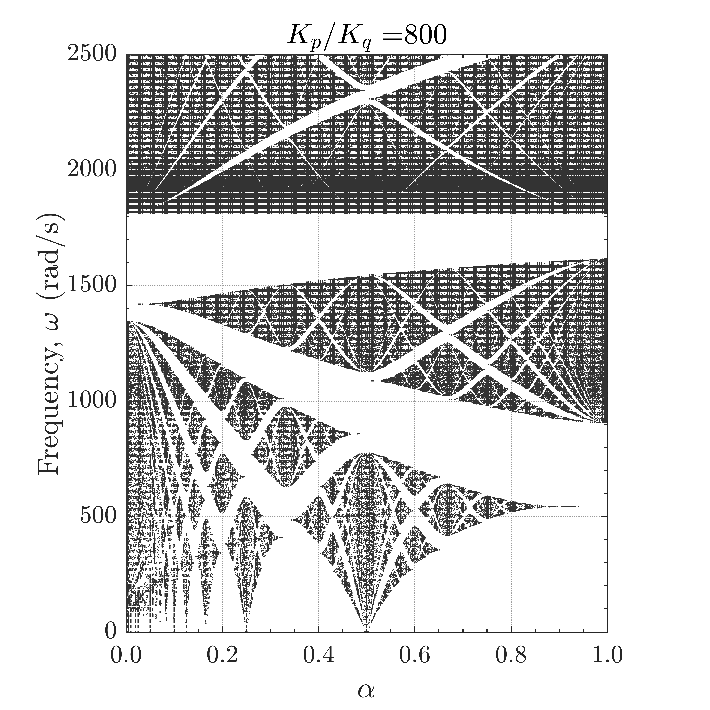}    			&			
			\includegraphics[width=8.5cm]{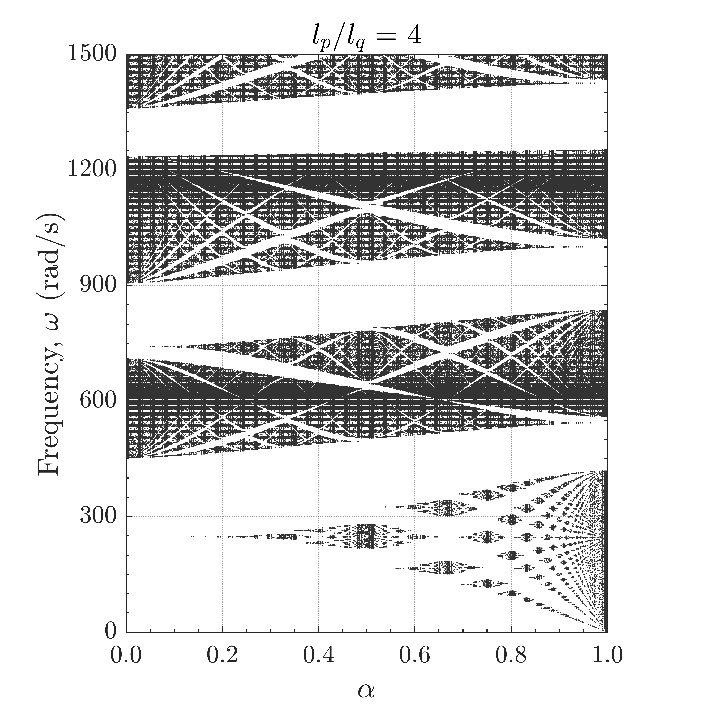}    			\\			
		\end{tabular}
		\caption{Bulk spectrum of structured Timoshenko beams using Sturmian quasiperiodic patterns. (a) Bending waves on a beam without supports with quasiperiodic distribution of the bending sectional stiffness, $EI$. Relationship between parameters in elements of type $p$ and $q$, $EI_p/EI_q = 30$. (b) Same case but with quasiperiodic distribution of the parameter mass per unit of length, $\rho A$, with ratio $\rho A_p / \rho A_q = 16$. (c) Continuous beam on elastic supports separated a distance $l=1$ m (constant) with quasiperiodic distribution of spring rigidities, $K_j \in \{K_p,K_q\}$ with $K_p/K_q = 800$. (d) Continuous beam on elastic supports with constant stiffness $K$. Separation pattern between  springs follows a quasiperiodic sequence with $l \in \{l_p, l_q\}$and $l_p / l_q = 4$.} %
		\label{fig12}%
	\end{center}
\end{figure}

Several numerical experiments have been carried out to show the variability of the pattern obtained by the Sturmian bulk spectrum. In Fig.~\ref{fig12} four cases have been represented. Each bulk spectrum has been plotted together with a sketch of the corresponding system. Thus, cases (a) and (b) are infinite unsupported beams where the variable parameters in the Sturmian sequences are the bending stiffness $EI$ and the mass per unit of length $\rho A$, respectively. Cases (c) and (d) are Timoshenko beams on elastic supports. In cases (c) and (d) the quasiperiodic parameters are respectively the rigidity $K_j$ and the distance $l_j$ between springs. The rest of the properties have been listed in the Table~\ref{Table_Properties_TIBeam}. \\

The case $EI_p/EI_q=30$ has been plotted in fig.~\ref{fig12}(a). It turns out that the square root of this ratio is not an integer number. Since the transfer matrix depends on expressions involving harmonic functions in terms of these parameters, then we can expect that in the final expression, although a closed form is not available, no periodicity will be shown in the frequency domain something that is revealed in figs.~\ref{fig12}(a). On the contrary, it has been imposed a ratio $\rho A_p / \rho A_q = 16 = 4^2$ (perfect square) something that leads to a periodic bulk spectrum in the vertical direction in fig.~\ref{fig12}(b).  Both cases (c) and (d) represent a beam on elastic supports of rigidity $K_j$ separated a distance $l_j$. In case (c) the distance $l_j = l_p = l_q$ remains constant and the spring constants are distributed in terms of the Sturmian sequence, i.e. $K_j \in \{K_p,K_q\}$. On the other side, in case (d), $K_j = K_p = K_q$ and the distance between consecutive springs alternates according to what is dictated by the corresponding quasiperiodic pattern, i.e. $l_j \in \{l_p,l_q\}$. The bulk spectrum of cases (c) and (d) reveals the effect of the distribution of the elastic supports in the low frequency range. Since the beam itself is continuous without heterogeneous distribution of parameters, it is expected that the high frequency range associated to short wavelength  will be less affected  by the stiffness of the supports. This behavior can be visualized in fig.~\ref{fig12}(c) and (d) where passbands are closer each other in the high frequency range.
\begin{table}
	\begin{tabular}{lccccccc}
 							&	$EI_j$ (Nm$^2$) &	$GA_j$ (N) &	$\rho A_j$ (kg/m)	&	$\rho I_j$ (m$^2$kg/m)	& $l_j$ (m)	& $K_j$ (N/m) \\
 							\hline 
Case \textbf{(a)}, $EI_p \neq EI_q$	
							&	0.2500/ 0.0083		&	3.00					&  0.010	&	8.33$\times 10^{-6}$		&	1.00  &	--- \\
Case \textbf{(b)}  $\rho A_p \neq \rho A_q $	
						    &	0.0083					  &	3.00					& 0.160/0.010	&	8.33$\times 10^{-6}$		&	1.00  &	--- \\
Case \textbf{(c)} 	$K_p \neq K_q$ 
							&	8.33	&	3.33 $\times 10^3$	&  0.010	&	8.33$\times 10^{-6}$ &	1.00  &	6.67$\times 10^3$/8.33   \\
Case \textbf{(d)} 	$l_p \neq l_q$ 
							&	8.33	&	3.33 $\times 10^3$					&  0.010	&	8.33$\times 10^{-6}$		&	4.00/1.00  &	6.67$\times 10^3$ \\
 							\hline 
\end{tabular}
	\caption{Example 3: Properties of Timoshenko beams in fig.~\ref{fig12}. 
			Each one of the four cases are associated with a quasiperiodic variation of a parameter. The two values of that parameter are indicated in the corresponding column as $(\bullet)_p/(\bullet)_q$}
	\label{Table_Properties_TIBeam}	
\end{table}
~\\

The Timoshenko beam model provides a glimpse into the range of possible bulk spectra as a function of the different parameters. Our motivation is still to find closed forms that allow predicting the general behavior of the Sturmian systems along the range $0 \leq \alpha \leq 1$, without the numerical computation of the dispersion relation for each $\alpha$, something that is computationally expensive, specially if $4 \times 4$ transfer matrices are involved. \\

    In reference to the self-similarity of the bulk-spectrum for a quasiperiodic Timoshenko beam, the developments presented in Sect. \ref{selfsimilarity} lead to the prediction that self-similar forms can be observed between two consecutive systems associated with parameters $\alpha = \{a,b\}$. This may give rise by repetition to visible fractal structures, as indeed can be observed in Figs. \ref{fig12}(a) to \ref{fig12}(d). The structure and organization of the system seems to greatly affect the spectrum, as indeed is seen in the four presented cases in the current example. For instance, Timoshenko beams in which sectional properties are quasiperiocally distributed (cases (a) and (b), see Table~\ref{Table_Properties_TIBeam}) seem to have a bulk-spectrum similar to those of rods (fig \ref{fig10}). However, the introduction of other elements, such as springs, introduce resonances that affect to the spectrum in the low frequency range, as explained above, although still showing signs of self-similarity, as can  be observed in figs \ref{fig12}(c) and \ref{fig12}(d). 
A deeper study of the modes of traveling waves in quasiperiodic Sturmian media and their relation to self-similarity may shed some light on this question, something that is currently under research. \\

The practical consequences of quasiperiodicity have been explored in the field of condensed matter, photonic and phononic quasicrystals, quasiperiodic dielectric multilayers, photovoltaic cells or even number recognition devices. Some of these real applications that take advantage of the fractal nature of quasiperiodic systems can be found in ref. Maci\'a \cite{Macia-2009}. Quasiperiodic structures, such as the ones studied in this work, are interesting because, if properly designed, they can cover large bandwidths. The fractal nature of the spectrum guaranties that you can find a quasiperiodic structure that has a bandgap placed in the frequency band of interest, structure identified by certain number $\alpha \in [0,1]$.  Nevertheless, in practical applications one must reach a balance between those effects related to the presence of energy losses and the dispersion effects (requiring relatively small systems) and beneficial aspects stemming from selfsimilarity and quasiperiodicity related effects (which require a large enough system).

\section{Conclusions}

In this paper we study the dynamic properties of heterogeneous elastic structured media with quasiperiodic pattern.  Sturmian words have been originally defined in the context of information theory. Quasiperiodic patterns based on Sturmian words have been applied in theoretical physics such as condensed matter or quantum mechanics, but to the best of our knowledge, they have not been applied to the generation of mechanical systems such as elastic waveguides. Consider any real number in the interval [0,1] given as a continued fraction, we can construct a word or sequence from a binary alphabet, giving rise to so-called Sturmian sequence. The methodology proposed in this paper takes such a string consisting of two symbols and transforms it into a quasi-periodic sequence consisting of two different values of one single parameter from the mechanical system. Dynamical properties and dispersion relations of Sturmian mechanical systems have been analytically determined using the transfer matrix method. This method also allows to justify self-similarity of the bulk spectrum. These properties have been validated and visualized along three numerical examples covering a  spring-mass lattice and two continuous systems: a rod and a beam. In the first case, the spring constants have been used as the quasiperiodic parameter. In the case of the rod, the sectional stiffness is changed according to the Sturmian pattern and in the beam case four different mechanical variables to be distributed quasiperiodically are considered separately: sectional stiffness, mass, spring supports rigidity and distances between springs. In all cases, the complete bulk spectrum of admitted states or frequencies of the system have been obtained and the results derived from the theoretical analysis validated. 

\section*{Acknowledgments}

This research was partially supported by the project with reference PID2020-112759GB-I00/AEI/10.13039/501100011033 from the Ministerio de Ciencia e Innovaci\'on (Spain) and by the project HYPERMETA funded under the program \'Etoiles Montantes of the R\'egion Pays de la Loire (France). The part of research of Agnieszka Niemczynowicz in this  publication  was written as a result of  internship in Valencia, Spain, co-financed by the European Union under the European Social Fund (Operational Program Knowledge Education Development), carried out in the project Development Program at the University of Warmia and Mazury in Olsztyn (POWR.03.05. 00-00-Z310/17).

\bibliography{bibliography}
\bibliographystyle{unsrt}
\end{document}